\documentclass{aa}  

\usepackage{graphicx,tablefootnote}
\usepackage{latexsym,amssymb}
\usepackage{amsmath}
\usepackage{gensymb}
\usepackage{capt-of}
\usepackage{textpos}
\usepackage{textcomp}
\usepackage{soul}
\usepackage{url}
\usepackage{multirow}
\usepackage{color}
\usepackage[T1]{fontenc}
\usepackage{longtable}
\usepackage{hyperref}
\usepackage{float}
\usepackage{siunitx}
\usepackage{textcomp}
\usepackage{txfonts}
\usepackage{xcolor}
\hypersetup{
    bookmarks=true,         % show bookmarks bar?
    unicode=false,          % non-Latin characters in Acrobat�s bookmarks
    pdftoolbar=true,        % show Acrobat�s toolbar?
    pdfmenubar=true,        % show Acrobat�s menu?
    pdffitwindow=false,     % window fit to page when opened
    pdfstartview={FitH},    % fits the width of the page to the window
    pdftitle={My title},    % title
    pdfauthor={Author},     % author
    pdfsubject={Subject},   % subject of the document
    pdfcreator={Creator},   % creator of the document
    pdfproducer={Producer}, % producer of the document
    pdfkeywords={keyword1} {key2} {key3}, % list of keywords
    pdfnewwindow=true,      % links in new window
    colorlinks=true,       % false: boxed links; true: colored links
    linkcolor=black,          % color of internal links (change box color with linkbordercolor)
    citecolor=black,        % color of links to bibliography
    filecolor=magenta,      % color of file links
    urlcolor=black,           % color of external links
    linkbordercolor={1 0 0},
    citebordercolor={0 1 0}
}
\makeatletter
\Hy@AtBeginDocument{%
  \def\@pdfborder{0 0 1}% Overrides border definition set with colorlinks=true
  \def\@pdfborderstyle{/S/U/W 0}% Overrides border style set with colorlinks=true
}
\makeatother
\newcommand{\kms}{km\,s$^{-1}$}

\newcommand{\Msun}{M$_\odot$}
\newcommand{\Lsun}{L$_\odot$}

\newcommand{\kmsM}{km\,s$^{-1}$\,Mpc$^{-1}$}      
      
\newcommand{\Reff}{\ensuremath{R_e}}

\newcommand{\ML}{\ensuremath{\Upsilon}}

\newcommand{\MLstar}{\ensuremath{\Upsilon_\star}}

\newcommand{\holi}{H0LiCOW}			     	%Project H0LiCOW
\newcommand{\hst}{\textit{HST}}             %HST
\newcommand{\jwst}{\textit{JWST}}           %JWST
\newcommand{\elt}{\textit{E-ELT}}           %E-ELT
\newcommand{\sersic}{S\'{e}rsic}		    %S�rsic
\newcommand{\lcdm}{$\Lambda$CDM}			%LCDM
\newcommand{\rxj}{RXJ1131$-$1231}             %WMAP
\newcommand{\h}{$H_\mathrm{0}$}				%H0
\newcommand{\kext}{\ensuremath{\kappa_\mathrm{ext}}}			%kext
\newcommand{\kint}{\ensuremath{\kappa_\mathrm{int}}}			%kint
\newcommand{\lint}{\ensuremath{\lambda_\mathrm{int}}}			%kint
\newcommand{\dt}{\ensuremath{D_{\Delta t}}}						%Ddt
\newcommand{\dtint}{\ensuremath{D_{{\Delta t},{\rm int}}}} 		%Ddt^mod
\newcommand{\dd}{\ensuremath{D_\mathrm{d}}}    					%Dd
\newcommand{\ddint}{\ensuremath{D_{\mathrm{d},{\rm int}}}}		%Dd,int
\newcommand{\ds}{\ensuremath{D_\mathrm{s}}}    					%Ds
\newcommand{\dds}{\ensuremath{D_\mathrm{ds}}}		    		%Dds
\newcommand{\zd}{\ensuremath{z_\mathrm{d}}}    					%Dds
\newcommand{\vlos}{\ensuremath{\overline{v_{\mathrm{LOS}}^{2}}}}%Vlos
\newcommand{\vlosinput}{\ensuremath{\overline{v_{\mathrm{LOS,\,input}}^{2}}}}%Vlosmock
\newcommand{\vlosinputl}{\ensuremath{\overline{v_{\mathrm{LOS,\,input,\,\textit{l}}}^{2}}}}%Vlosmockl
\newcommand{\vlosldata}{\ensuremath{\overline{v_{\mathrm{LOS,\,data,\,\textit{l}}}^{2}}}}%Vlosdatal
\newcommand{\vlosmed}{\ensuremath{\overline{v_{\mathrm{LOS,\,med}}^{2}}}}%Vlosmed
\begin{document}

   \title{TDCOSMO VIII: Cosmological distance measurements in light of the mass-sheet degeneracy - forecasts from strong lensing and IFU stellar kinematics}

%   \subtitle{}

   \author{A. Y{\i}ld{\i}r{\i}m
   \inst{1},
   S.~H.~Suyu,
   \inst{1,2,3}, 
   G. C.-F.~Chen,
   \inst{4}
   E. Komatsu
   \inst{1,5}}

   \institute{Max Planck Institute for Astrophysics, Karl-Schwarzschild-Str. 1, 85748 Garching, Germany
   \email{yildirim@mpa-garching.mpg.de}
   \and
   Technische Universit\"at M\"unchen, Physik-Department, James-Franck-Str. 1, 85748 Garching, Germany
   \and
   Institute of Astronomy and Astrophysics, Academia Sinica, 11F of ASMAB, No.1, Section 4, Roosevelt Road, Taipei 10617, Taiwan
   \and
   Department of Physics and Astronomy, University of California, Los Angeles, CA 90095, USA
   \and
   Kavli Institute for the Physics and Mathematics of the Universe (Kavli IPMU, WPI), UTIAS, The University of Tokyo, Chiba, 277-8583, Japan}

%   \date{Received September 15, 1996; accepted March 16, 1997}

% \abstract{}{}{}{}{} 
% 5 {} token are mandatory
 
  \abstract
  {Time-delay distance measurements of strongly lensed quasars have provided a powerful and independent probe of the current expansion rate of the Universe (\h). However, in light of the discrepancies between early and late-time cosmological studies, current efforts revolve around the characterisation of systematic uncertainties in the methods. In this work, we focus on the mass-sheet degeneracy (MSD), which is commonly considered as being a significant source of systematics in time-delay strong lensing studies, and aim to assess the constraining power provided by integral field unit (IFU) stellar kinematics. To this end, we approximate the MSD with a cored, two-parameter extension to the adopted lensing mass profiles (with core radius $r_{\rm c}$ and mass-sheet parameter \lint), which introduces a full degeneracy between \lint\ and \h\ from lensing data alone. In addition, we utilise spatially resolved mock IFU stellar kinematics of time-delay strong lenses, given the prospects of obtaining such high-quality data with the James Webb Space Telescope (\jwst) in the near future. We construct joint strong lensing and generalised two-integral axisymmetric Jeans models, where the time delays, mock imaging and IFU observations are used as input to constrain the mass profile of lens galaxies at the individual galaxy level, and consequently yield joint constraints on the time-delay distance (\dt) and angular diameter distance (\dd) to the lens. We find that mock \jwst-like stellar kinematics constrain the amount of \textit{internal} mass sheet that is physically associated with the lens galaxy and limit its contribution to the uncertainties of \dt\ and \dd, each at the $\le$ 4\% level, without assumptions on the background cosmological model. Incorporating additional uncertainties due to external mass sheets associated with mass structures along the lens line-of-sight, these distance constraints would translate to a $\lesssim$ 4\% precision measurement on \h\ in flat \lcdm\ cosmology for a \textit{single} lens. Our study shows that future IFU stellar kinematics of time-delay lenses will be key in lifting the MSD on a per lens basis, assuming reasonable and physically motivated core sizes. However, even in the limit of infinite $r_{\rm c}$, where \dt\ is fully degenerate with \lint\ and is thus not constrained, stellar kinematics of the deflector, time delays and imaging data will provide powerful constraints on \dd, which becomes the dominant source of information in the cosmological inference.}

   \titlerunning{Cosmological distance forecasts with MSD}

   \maketitle
\section{Introduction}
\label{sec:introduction}
%=====================================================================

The current expansion rate of the Universe (\h) is a fiercely debated issue. By means of many independent probes, and efforts spanning multiple decades, a picture has now emerged in which there appears to be a clear discrepancy between early and late-time cosmological studies \citep{2020A&A...641A...6P,2019ApJ...876...85R,2020MNRAS.498.1420W,2018MNRAS.480.3879A,2020ApJ...891L...1P,2019NatAs...3..891V}. At the heart of this debate is nothing less than our understanding of our standard cosmological model and, at face value, this discrepancy can only be resolved by yet unknown and unaccounted physics \citep[e.g.,][]{2019PhRvL.122v1301P,2019arXiv190401016A,2020PhRvD.101d3533K} or by our lack of accounting for systematic uncertainties in the methods \citep[e.g.,][]{2019ApJ...882...34F,2019MNRAS.484L..64S}.  %SHS: there are >2 methods, and either refers to only 2 methods
 
Given the possibly drastic implications, the assessment of the systematics has been of particular interest lately. In the context of time-delay strong lensing \citep[TDSL;][]{1964MNRAS.128..307R,2016A&ARv..24...11T,2018SSRv..214...91S}, the prominent mass-sheet degeneracy\footnote{A special case of the Source-Position Transformation \citep{2014A&A...564A.103S,2017A&A...601A..77U,2018A&A...617A.140W}} \citep[MSD;][]{1985ApJ...289L...1F} is commonly quoted as being a fundamental source of uncertainty \citep{2013A&A...559A..37S,2020MNRAS.493.1725K}. In brief, the MSD manifests itself through a degeneracy between the lens potential and the observationally inaccessible source position and size. That is, a transformation of the former can be compensated by a translation and scaling of the latter \citep[see Eq. 6-12 in][and references therein]{2020MNRAS.493.4783Y}. This mass-sheet transformation (MST) leaves many observables invariant and thus recovers e.g. the image positions and time delays equally well. However, with the time-delay distance, and hence \h, being sensitive to the (functional/radial form of the) lens potential \citep{2018MNRAS.474.4648S}, accounting for the MSD is crucial for obtaining an unbiased measurement of \h\ with realistic uncertainties.

To reduce the complexity of the MSD, we follow the classification introduced in \cite{2010ApJ...711..201S} and differentiate between two kinds of mass sheets: i) an \textit{internal} mass sheet that is physically associated with the strong lens galaxy and hence affects its stellar kinematics, and ii) an  \textit{external} mass sheet that is attributed to mass structures along the strong-lens line-of-sight (LOS) which are not physically associated with the strong lens galaxy and thus do not affect its stellar kinematics.
The \holi\ \citep[\h\ Lenses in COSMOGRAIL's\footnote{COSmological MOnitoring of GRAvItational Lenses; \url{https://cosmograil.epfl.ch}} Wellspring;][]{2017MNRAS.468.2590S} and TDCOSMO\footnote{\url{http://www.tdcosmo.org/}} \citep{2020A&A...639A.101M} collaborations account for \textit{external} mass sheets by explicitly modelling e.g. the nearest and brightest galaxies (in projection) along the LOS and comparing observed galaxy counts in the vicinity of the strong-lens system to counts from ray tracing through simulations. These efforts show that a final precision of 2.4\% on \h\ is achieved from an ensemble of 6 time-delay strong lenses, when well motivated mass models for the lens galaxies are adopted \citep{2020MNRAS.498.1420W}. However, if these mass model assumptions are relaxed and a model with \textit{internal} mass sheet -- that is maximally degenerate with \h\ -- is used, then the constraint on \h\ degrades to 5\% or 8\%, respectively, depending on whether or not external priors (obtained from non-time-delay lenses) for the orbital anisotropy are adopted \citep{2020A&A...643A.165B}. 
%Owing to this relaxation of the lens mass profile, the most recently reported value of \h$=67.4^{+4.1}_{-3.2}$\,\kmsM\ from TDCOSMO is now statistically in agreement with measurements from the Planck collaboration. 

In order to gain back the precision lost by this more general and inclusive approach, several avenues can be explored. These range from acquiring even more stellar kinematic data of non-time-delay lenses (to further constrain the orbital anisotropy and mass profile of lens galaxies at the galaxy population level) to high-quality, partially resolved stellar kinematics (i.e., kinematics within azimuthally averaged rings from the ground) of individual time-delay lenses \citep{2021A&A...649A..61B}.

%Indeed, this inherent modelling limitation has been picked up by \citet[][herafter B20]{2020ApJ...892L..27B} recently, where a family of mass models has been introduced to alleviate the apparent discrepancy in the \h\ constraints reported by the \holi\ and Planck Collaboration. By means of an inner cored density distribution, on top of a broader power-law profile, B20 are capable of recovering both the image position and relative time delays of mock strong lens systems. This cored density distribution is highly degenerate with a power-law lens mass model - commonly employed in strong lensing studies \citep{2006ApJ...649..599K} - and thereby allows for a shift in \h, depending on its contribution to the total mass profile. Hence, such a cored density profile, presumably composed of non baryonic matter, would allow for a realignment of TDSL and CMB results.
%
%While the proposed family of mass models in B20 presents an innovative approach at solving the \h\ controversy, it fails to provide a physical explanation for the presence of such a cored density distribution and the lack of observational evidence.

Given our previous work on forecasting cosmological distance measurements with high-spatially resolved stellar kinematics from the next generation telescopes, such as the James Webb Space Telescope (\jwst) or the planned European Extremely Large Telescope (\elt) \citep[][hereafter Y20]{2020MNRAS.493.4783Y}, this study aims to reassess the constraining power of joint strong lensing and stellar dynamical models for cosmological purposes, while incorporating now a more general family of mass models that closely mimics the contribution of an \textit{internal} mass-sheet component. The mass-sheet component used in this work resembles the parameterisation presented in \citet[][herafter B20]{2020ApJ...892L..27B} of a cored mass-density distribution, which has been tailored to be insensitive to the lensing data and is therefore maximally degenerate with \h. Adopting this cored density distribution hence allows us to fully explore the uncertainties associated with the MSD and the prospects of limiting its impact effectively by means of integral field unit (IFU) data, which is otherwise not possible unless, e.g., aided by additional distance indicators, the assumption of a cosmological world model \citep{2020arXiv201106002C} or the aforementioned use of prior information \citep{2020A&A...643A.165B}.

Throughout this paper, we adopt a standard cosmological model with \h\ = 82.5 \kmsM, a matter density of $\Omega_{m} = 0.27$, and a dark energy density of $\Omega_{\Lambda} = 0.73$, where our particular choice for \h\ is driven by the time-delay distance measurement of \rxj\ in \cite{2014ApJ...788L..35S}

%============================= Section 2 =============================
\section{Data}
\label{sec:data}
%=====================================================================

\begin{figure}
\begin{center}
\includegraphics[width=0.9\linewidth]{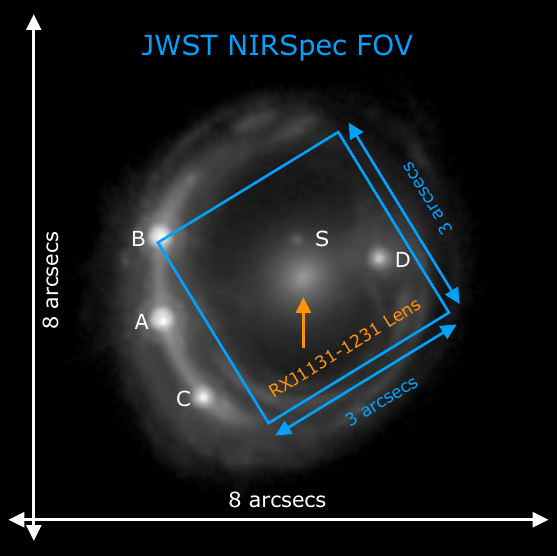}
\caption{\hst\ ACS F814W imaging cutout of \rxj, illustrating the prominent lens configuration with a quadruply imaged background quasar (A, B, C \& D) and a nearby satellite (S). Spectroscopic measurements locate the lens and quasar at redshifts $z_{\mathrm{d}} = 0.295$ and $z_{\mathrm{s}} = 0.654$, respectively. We generated mock imaging data, closely resembling the archival \hst\ data shown here, based on a best-fitting lensing model and excluding the satellite. Overlaid is the \jwst\ NIRSpec nominal FOV of 3\arcsec$\times$3\arcsec, within which we create mock stellar kinematics of the foreground lens at 0.1\arcsec/pixel resolution. The FOV is oriented such that the x-axis is aligned with the galaxy major axis.}
\label{fig:fig1}
\end{center}
\end{figure}

As a test-bed for our study, we will make use of \rxj\ that was discovered by \citet{2003A&A...406L..43S}. Being part of the \holi\ base sample \citep{2017MNRAS.468.2590S}, \rxj\ represents one of the most prominent TDCOSMO systems with exquisite data, consisting of Hubble Space Telescope (\hst) imaging (Fig. \ref{fig:fig1}) and highly precise time-delay measurements \citep{2013A&A...553A.120T,2020A&A...640A.105M}. In addition, an aperture averaged stellar velocity dispersion measurement for the lensing galaxy of $\sigma = 323\pm20$ \kms\ is available \citep{2013ApJ...766...70S}. 

Following the procedure outlined in Y20, we make use of a hybrid data strategy, where the archival \hst\ data (GO: 9744; PI: Kochanek) and literature time-delays \citep{2013A&A...556A..22T} are aided by mock IFU stellar kinematics. In detail, the \hst\ observations have been used to remock the imaging data. This is achieved by means of the best-fitting \texttt{COMPOSITE} (i.e. stars \& NFW halo) lensing model achieved on the \hst\ lens image (Fig. \ref{fig:fig1}), as obtained in \cite{2014ApJ...788L..35S}. In this model, the extended source has been reconstructed on a source grid with a resolution of $64 \times 64$ pixels. Unlike the \texttt{COMPOSITE} lens model in \cite{2014ApJ...788L..35S}, we excluded the nearby satellite (S in Fig. \ref{fig:fig1}) in our mock \hst\ image, to ensure consistency with the mock IFU stellar kinematics, where the satellite has also been omitted (see Sec. \ref{sec:framework} for further explanation). The stellar kinematics of \rxj\ have been generated based on this \texttt{COMPOSITE} lens mass model and using arbitrary dynamical model parameters (see Sec. \ref{sec:framework}). Utilising the axisymmetric Jeans formalism for mocking up the stellar kinematics \citep{1987gady.book.....B,2008MNRAS.390...71C,2012MNRAS.423.1073B}, we obtain IFU data at \jwst\ resolution (i.e. 0.1\arcsec/pixel), covering a 3\arcsec$\times$3\arcsec\ field-of-view (FOV). The mock IFU data closely resemble the Near-Infrared Spectrograph (NIRSPEC) properties and characteristics, assuming that the kinematics have been measured with a single pointing centred on \rxj\ (Fig. 1). 

The mock kinematics are further convolved with a single Gaussian PSF of 0.08\arcsec\ FWHM, which is already two times larger than the diffraction limit of \jwst. Difficulties in properly measuring the PSF size have been shown to be of minimal concern (Y20), which is why this PSF is employed for both the mock and modelling stage.

Since the signal-to-noise (S/N) in each spatial resolution element is usually insufficient to reliably measure the stellar kinematics across the entire FOV, we perform a spatial binning of the mock data \citep{2003MNRAS.342..345C}. To this end, the surface brightness (SB) distribution at \jwst\ resolution is obtained from parameterised fits to the \hst\ imaging data. This SB distribution of \rxj\ is then used to achieve a desired S/N level, by means of \jwst's exposure time calculator (ETC V1.3)  Within the 3\arcsec$\times$3\arcsec\ FOV, this S/N map is, in turn, used to spatially bin the individual pixels to a target S/N of $\sim$ 40 per bin. Our procedure results in $\sim$ 80 spatially resolved measurements of the line-of-sight velocity distribution (LOSVD) across the entire FOV, achievable with reasonable on-source integration times of 8-10h.

To imitate realistic observations, various sources of statistic and systematic uncertainties %correlated and uncorrelated errors 
are added to the mock input kinematics $\bigg(\sqrt{\vlosinput}\bigg)$. Here, we focus on two individual data sets. Following the notation in Y20, the IDEAL data incorporate random Gaussian errors only ($\delta v_{\rm{stat}}$). More specifically, the error in each bin $l$ is drawn from a Gaussian distribution with a mean of zero and a standard deviation ($\sigma_{\rm{stat,\,\textit{l}}}$) that is inversely proportional to its signal-to-noise (S/N). Given that all bins reach the target S/N of 40, this approach yields $\sigma_{\rm{stat,\,\textit{l}}} \approx 0.025 \times\ \sqrt{\vlosinputl}$. The IDEAL data hence reflect a best-case scenario for breaking the MSD, where the results will be unaffected by e.g. systematic effects in the measurement of the stellar kinematics. In the second data set, which we label as FIDUCIAL, we incorporate systematic errors in addition to the random Gaussian noise in the IDEAL data set. That is, the FIDUCIAL data set contains a 2\%\ systematic shift downward
%to the downside 
of all mock $\sqrt{\vlosinputl}$ values $\bigg(\delta v_{\rm{sys}} = -0.02 \times \sqrt{\vlosmed}\bigg)$, where $\bigg(\sqrt{\vlosmed}\bigg)$ is the median of $\sqrt{\vlosinputl}$ from all bins.
This systematic shift to the downside is introduced to account for possible systematics in the measurement of the stellar kinematics (e.g. due to stellar template mismatches). The mock kinematics can be quickly summarised as follows:

\begin{equation}
\sqrt{\vlosldata} = \underbrace{\overbrace{\sqrt{\vlosinputl} + \delta v_{\rm{stat}}}^{\rm{\textcolor{orange}{IDEAL}}} + \delta v_{\rm{sys}}}_{\rm{\textcolor{blue}{FIDUCIAL}}}
\label{eqn:eqn1}
\end{equation}

To check for consistency with real observations, we have collapsed the mock IFU data within an 0.8\arcsec\ wide aperture for the IDEAL case, which yields a mock single second-order velocity moment along the LOS of $\sqrt{\ensuremath{\overline{v_{\mathrm{LOS,ap}}^{2}}}} = 325\pm12$ \kms. This mock value is in excellent agreement with the literature stellar velocity dispersion measurement, assuming that rotation is minimal, and confirms the validity of our approach.

The mock \hst\ observations, time-delay measurements and mock kinematics are then employed as constraints for our joint modelling machinery and the modelling results for both mock data sets (i.e. IDEAL and FIDUCIAL) are presented and discussed in Sec. \ref{sec:framework} and \ref{sec:results}.

%============================= Section 3 =============================
\section{Theoretical Framework}
\label{sec:framework}
%=====================================================================
\subsection{Strong lensing under the internal MST}
\label{subsec:stronglensing}

We adopt the \texttt{COMPOSITE} mass model of \rxj\ as our baseline model (Y20). This model consists of a baryonic and non-baryonic component. The former is parameterised by multiple pseudo-isothermal elliptical profiles \citep[\texttt{PIEMDs};][]{Kassiola+93}, to resemble a two-component \sersic\ distribution, multiplied with a spatially constant stellar mass-to-light ratio (\ML), and the latter by a NFW profile \citep{1996ApJ...462..563N,1997ApJ...490..493N}. Moreover, we add a cored density distribution to the \texttt{COMPOSITE} model, here also parameterised as a \texttt{PIEMD}, with a core radius of $\theta_{\rm c} = 20\arcsec$, which is well beyond the Einstein radius ($\theta_{\rm E} \approx 1.6\arcsec$)\footnote{We refer to $\theta_{\rm c}$ as the angular core radius. The physical core radius $r_{\rm c}$ in units of kpc is obtained via $r_{\rm c} = \theta_{\rm c}\,\dd$. See also Sec. \ref{eqn:AppC:SMD}}. As introduced in B20, such a cored density distribution is built to mimic an \textit{internal} mass sheet with constant convergence $\kint$. In concordance with earlier notations, we define
\begin{equation}
%SHS: \kappa_{\rm int} = \lambda,
\lint \equiv 1 - \kint,
\label{eqn:eqn2}
\end{equation}
with
\begin{equation}
%SHS: \kappa_{\rm int} = \lambda,
\lint  \in [0.5, 1.5].
\label{eqn:eqn3}
\end{equation}
%SHS: with $\lambda \in [-0.5, 0.5]$.
By scaling the total convergence profile according to 
\begin{equation}
%SHS: \kappa_{\lambda}(\theta) = \lambda+(1-\lambda)\kappa_{\rm composite}(\theta),
\kappa_{\rm mst,int}(\theta) = (1-\lint) + \lint\;\kappa_{\rm composite}(\theta),
\label{eqn:eqn4}
\end{equation}
where $\kappa_{\rm composite}(\theta)$ is the original \texttt{COMPOSITE} profile, this MST leaves all observables invariant, provided the time-delay distance (\dt)
\begin{equation}
\dt  \equiv (1+\zd)\;\frac{\dd\ds}{\dds} \propto H_0^{-1},
\label{eqn:eqn5}
\end{equation}
scales as
\begin{equation}
%SHS: {D_{\Delta t, \lambda}} = \frac{\dtmod}{(1-\kappa_{\rm int})}.
\dtint = \frac{{D_{\Delta t, {\rm composite}}}}{\lint}.
\label{eqn:eqn6}
\end{equation}
Here $\zd$ is the redshift to the lens, $\dd$, $\ds$ and $\dds$ are angular diameter distances to the lens, to the source and between the lens and the source, respectively.
Incorporating further the effect of an \textit{external} mass-sheet (\kext) yields
\begin{equation}
\dt = \frac{\dtint}{1-\kext},
\label{eqn:eqn7}
\end{equation}
whereas \dd\ is fully invariant under the \textit{external} MST\footnote{The angular diameter distance to the lens \dd\ is also approximately invariant under the \textit{internal} MST with a single aperture averaged velocity dispersion and for \lint\ $\in [0.8,1.2] $\citep{2020arXiv201106002C}.} \citep{2015JCAP...11..033J} 
\begin{equation}
\label{eqn:eqn8}
\dd = \ddint.    
\end{equation}

In concordance with B20, our adopted cored density distribution has the special attribute of leaving the mean convergence within $\theta_{\rm E}$ ($\sim$ 1.6\arcsec) virtually identical and making it thus inaccessible for strong lensing studies alone. However, the above transformation introduces changes in the radial 3D mass distribution, depending on the extent of the core radius, to which spatially resolved stellar kinematics can be sensitive. Here, the choice of a core radius is particularly important. Our core radius of 20\arcsec\ might seem deliberate at first, but was adopted after intensive testing. Larger core radii will mimic a scenario in which the original density profile of the galaxy ($\kappa_{\rm composite}$) is effectively unchanged (see Appendix \ref{sec:appendixc}), i.e. both the lensing and kinematic data will be insensitive, whereas core radii which are too small will have a more noticeable imprint on the stellar kinematics, but might not be a \textit{valid} mass-sheet transformation in the sense that the lensing data will start to become sensitive to such a transformation of the mass profile \citep[see also Fig.3 in,][]{2020A&A...643A.165B}. The core radius of 20\arcsec\ was chosen to probe a worst-case scenario of sorts, where the MST is still intact from a lensing point of view while the imprint on the kinematic data is big enough to allow for meaningful constraints on $\lint$. Our distinction between an \textit{internal} and \textit{external} mass-sheet is therefore a practical one; we assume that the \textit{internal} MST does not conspire in a way that it becomes undetectable for both lensing and stellar kinematics within $\theta_{\rm E}$. We also note that the adopted core radius of 20\arcsec\ corresponds to a physical radius of $\approx 80$\,kpc at our lens distance. It is therefore sufficiently large to test physically motivated mechanisms that could produce extended cored density distributions, such as those put forward by ultralight dark matter \citep[see e.g.,][]{2021arXiv210510873B}.

\subsection{Stellar dynamics under the internal MST}
\label{subsec:stellardynamics}
We briefly revisit the theoretical framework for mocking up and modelling the stellar kinematics. As outlined in Y20, our models are based on the solution of the axisymmetric Jeans equations \citep{1922MNRAS..82..122J,1987gady.book.....B}, assuming alignment of the velocity ellipsoid with the cylindrical coordinate ($R, \phi, z$) system as well as a flattening in the meridional plane \citep{2007MNRAS.379..418C,2008MNRAS.390...71C}. This yields the two equations,
\begin{equation}
\frac{\beta_{z}\rho_*\overline{v_{z}^{2}}-\rho_*\overline{v_{\phi}^{2}}}{R}+\frac{\partial(\beta_{z}\rho_*\overline{v_{z}^2})}{\partial R} = -\rho_*\frac{\partial\psi_{\mathrm{D}}}{\partial R}
\label{eqn:eqn9}
\end{equation}
\begin{equation}
\frac{\partial(\rho_*\overline{v_{z}^{2}})}{\partial z} = -\rho_*\frac{\partial\psi_{\mathrm{D}}}{\partial z}
\label{eqn:eqn10}
\end{equation}
which link the tracer density  ($\rho_*$) and gravitational potential ($\psi_{D}$) to three intrinsic second-order velocity moments ($\overline{v_{z}^{2}}, \overline{v_{\phi}^{2}}, \overline{v_{R}^{2}}$). Once the tracer density is measured (by means of the SB distribution) and a gravitational potential is adopted, we can readily solve Eq. \ref{eqn:eqn10} for $\overline{v_{z}^{2}}$ and Eq. \ref{eqn:eqn9} for $\overline{v_{\phi}^{2}}$ and obtain $\overline{v_{R}^{2}}$ via $\beta_{z} = 1-\overline{v_{z}^{2}}/\overline{v_{R}^{2}}$. The intrinsic second-order moments can then be projected along the LOS to predict a second-order velocity moment
\begin{equation}
\begin{split}
\overline{v_{\mathrm{LOS}}^{2}} = &\frac{1}{\mu(x',y')}\;\int_{-\infty}^{\infty}\;\rho_*\Big[(\overline{v_{R}^{2}}\sin^{2}\phi+\overline{v_{\phi}^{2}}\cos^{2}\phi)\sin^{2}i \\
& +\overline{v_{z}^{2}}\cos^{2}i-\overline{v_{R}v_{z}}\sin\phi\sin(2i)\Big]dz'\;\equiv\;v^{2}+\sigma^{2},
\end{split}
\label{eqn:eqn11}
\end{equation}
where $x'$ and $y'$ are the cartesian coordinates on the plane of the sky, $z'$ is the coordinate along the LOS, $i$ is the inclination angle (with 90\degree\ being edge-on), $\mu$ is the observed SB and $v$ and $\sigma$ the observed mean LOS velocity and velocity dispersion.

In our self-consistent modelling approach, both the intrinsic tracer and mass density for the dynamics are constructed by deprojecting the SB and surface mass density (SMD) distribution from the lens models, which are constrained by the quasar image positions, time delays and the spatially extended Einstein ring. As mentioned in Sec. \ref{subsec:stronglensing}, the projected SMD consists of a \texttt{COMPOSITE} and \texttt{PIEMD} profile and is subject to the \textit{internal} MST. More specifically,
\begin{equation}
\Sigma_{\rm smd}(\theta) = \Sigma_{\rm crit} \times\ \kappa_{\rm mst, int}(\theta),
\label{eqn:eqn12}
\end{equation}
where $\Sigma_{\rm smd}$ denotes the (projected) physical surface mass density and
\begin{equation}
\Sigma_{\mathrm{crit}} = \frac{c^{2}}{4\pi G}\;\frac{\dtint}{(1+\zd)\;\ddint}\;\frac{1}{\ddint}
\label{eqn:eqn13}
\end{equation}
is the critical surface mass density\footnote{We can express $\Sigma_{\mathrm{crit}}$ here in terms of $\dtint$ and $\ddint$, instead of $\dt$ and $\dd$, because a subsequent external MST (after the internal MST) does not affect $\Sigma_{\rm smd}$ as the terms connected to $\kext$ cancel out (Y20).}.
With Eq. \ref{eqn:eqn4}, we rewrite the projected SMD as
\begin{equation}
\begin{aligned}
\Sigma_{\rm smd}(\theta) & = \bigg[ \frac{c^{2}}{4\pi G}\;\frac{{\dtint}}{(1+\zd)\;\ddint}\;\frac{1}{\ddint}\bigg]\\
& \times \bigg[(1-\lint)+\lint\;\kappa_{\rm composite}(\theta) \bigg].\\
\end{aligned}
\label{eqn:eqn14}
\end{equation}
The SMD is expanded by multiple Gaussians \citep[MGE;][]{1994A&A...285..723E} and deprojected into a 3D mass density. Assuming an oblate axisymmetric system, the intrinsic total density of each Gaussian in the fit \citep[MGEfit;][]{2002MNRAS.333..400C} is obtained via
\begin{equation}
\rho_{i}(R, z) = \frac{M_{i}}{(\sigma_{i} \sqrt{2 \pi})^3\;q_{i}}\;\exp\bigg[-\frac{1}{2\sigma_{i}^{2}}\;\Bigg(R^{2} + \frac{z^2}{q_{i}^2} \Bigg)\bigg],
\label{eqn:eqn15}
\end{equation}
and related to $\psi_{\rm D}$ through Poisson's equation. We use index $i$ to denote the MGE components of the mass density. It is of importance to note that, based on our finite core size, our treatment and implementation of the MST is an approximate one, with $\kint \approx \mathrm{const.}$ within $\theta_{\rm c}$. This, however, is a good enough approximation of the perfect MST (i.e. $\theta_{\rm c} \to \infty$), where the deviation in e.g. the deflection angles are small enough (i.e. $\delta_{\theta}(\theta_{\rm c} \to \infty) \approx\ \delta_{\theta}(\theta_{\rm c} = 20\arcsec)$), such that the models fully capture the degeneracies associated with it (see Sec \ref{sec:results}).

In Eq. \ref{eqn:eqn15}, $M_{i} = 2\,\pi\,\sigma_{i}^{'2}\,q'_{i}\,\Sigma_{\mathrm{M},i}$ denotes the total mass of the individual Gaussians, $\sigma'_{i}$ the projected dispersion, $q'_{i}$ the projected flattening and $\Sigma_{\mathrm{M},i}$ the peak surface mass density. In our case of a MGE of the SMD, the gravitational potential assigned to each Gaussian is
\begin{equation}
\psi_{{\rm D},i}(R,z) = -\frac{2G\,M_{i}}{\sqrt{2\pi}\sigma_{i}}\;\int_{0}^{1}\;\mathcal{F}_{i}(u)du
\label{eqn:eqn16}
\end{equation}
with
\begin{equation}
\mathcal{F}_{i}(u) = \exp \bigg[-\frac{u^{2}}{2\sigma^{2}_{i}}\Bigg(R^{2}+\frac{z^2}{\mathcal{Q}^{2}_{i}(u)}\Bigg)\bigg]\frac{1}{\mathcal{Q}_{i}(u)}
\label{eqn:eqn17}
\end{equation}
and $\mathcal{Q}^{2}_{i}(u) = 1-(1-q^{2}_{i})\;u^{2}$.
Here, $\sigma_{i}$ is the intrinsic dispersion of the Gaussians, $q_{i}$ their intrinsic long vs. short axis ratio, and ($R$, $z$) the intrinsic cylindrical coordinate system.
Similarly, the SB distribution can also be expanded by multiple Gaussians.
Denoting the MGE components of the luminosity distribution by index $j$ (in contrast to $i$ for the mass density),
%and 
the intrinsic luminosity density of each Gaussian is
\begin{equation}
\rho_{*,j}(R, z) = \frac{L_{j}}{(\sigma_{j} \sqrt{2 \pi})^3\;q_{j}}\;\exp\bigg[-\frac{1}{2\sigma_{j}^{2}}\;\Bigg(R^{2} + \frac{z^2}{q_{j}^2} \Bigg)\bigg],
\label{eqn:eqn18}
\end{equation}
with $L_{j}$ now expressing the total luminosity of the individual Gaussians. In general, the set of Gaussians describing the SB and SMD are not identical (i.e. $i \neq j$ and hence $\sigma_{i} \neq \sigma_{j}$ and $q_{i} \neq q_{j}$) unless mass-follows-light. 
The second-order moments of a single Gaussian tracer distribution -- embedded in the gravitational potential $\psi_{\rm D} = \sum_{i} \psi_{{\rm D},i}$ -- is then obtained via
\begin{equation}
[\overline{v_{z}^{2}}]_{j} = 4\pi\,G \sum_{i} \frac{\Sigma_{\mathrm{M}, i}\,q'_{i}}{\sqrt{2\pi}\sigma_{i}}\int_{0}^{1}q_{j}^{2}\sigma_{j}^{2}\frac{\mathcal{F}_{i}(u)u^{2}}{1-\mathcal{F}_{ji}u^{2}}du,
\label{eqn:eqn19}
\end{equation}
with $\mathcal{F}_{ji} = 1 - q_{i}^{2} - q_{j}^{2}\sigma_{j}^{2}/\sigma_{i}$ \citep[see][for analytical expressions of all velocity moments involved in Eq. \ref{eqn:eqn9} and \ref{eqn:eqn10}]{2008MNRAS.390...71C,2010ApJ...719.1481V}. Note that absolute lengths in the plane of the sky scale linearly with the distance ($r \propto \ddint$). Hence, when moving from angular to physical units, this dependency translates into
\begin{equation}
    \vlos \propto 1/\ddint.
\label{eqn:eqn20}
\end{equation}

\begin{figure}
\begin{center}
\includegraphics[width=1\linewidth]{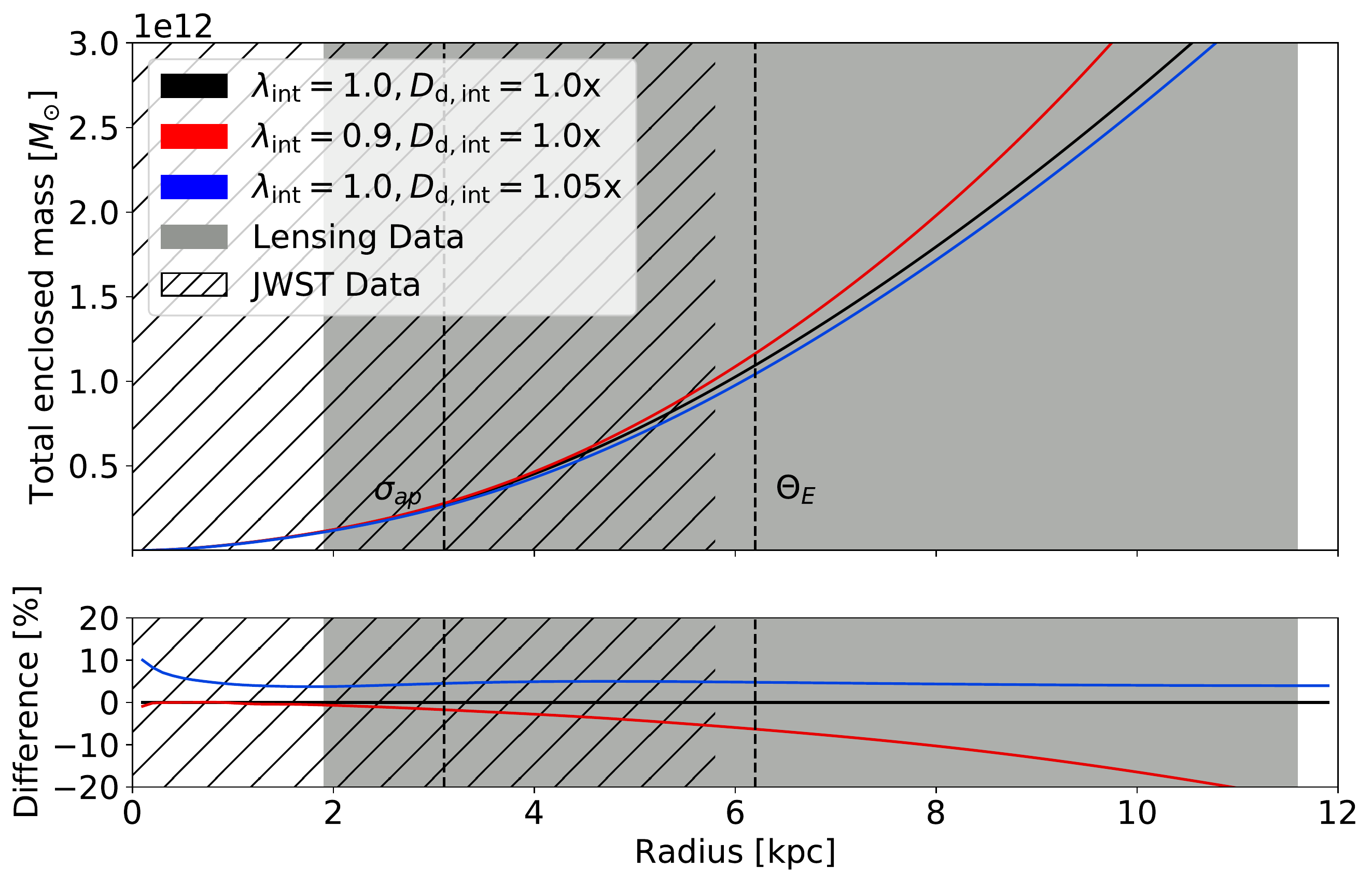}
\caption{Top: total enclosed mass profile in 3D as a function of intrinsic radius ($\sqrt{R^{2}+z^{2}}$), after deprojecting the SMD into an oblate axisymmetric system. The black line shows the mock input mass profile (i.e. the \texttt{COMPOSITE} model without a mass sheet) and the red line shows the mass-sheet transformed profile with $\lint = 0.9$. The profile without a mass-sheet transformation, but with \ddint\ scaled by +5\%, is shown in blue. The vertical dashed lines indicate the extent of literature, aperture averaged measurement of the velocity dispersion ($\sigma_{\rm ap}$) and the Einstein radius ($\theta_{\rm E}$) in projection. The grey shaded area shows the region where the lensed imaging data constrain the mass model and the striped area roughly corresponds to \jwst's FOV. Bottom: the percent differences in the total enclosed mass as a function of radius when compared to the mock input mass profile (black). The figure illustrates that a single aperture averaged velocity dispersion is insufficient for constraining the mass sheet, given the minuscule differences in the total enclosed mass within that radius and errors on the kinematics of the order of 5\%. On the other hand, spatially resolved kinematics are sensitive to both, differences in the enclosed mass as well as the mass slope. As is evident from the figure, \ddint\ and \lint\ are not degenerate. That is, changes in \lint\ (and hence \dtint) cannot be compensated by a corresponding scaling of the lens distance \ddint. Numerical noise from the MGE is present, but its contribution to the observed differences is subdominant with respect to the final modelling uncertainties.}
\label{fig:fig2}
\end{center}
\end{figure}

It is immediately evident from Eq. \ref{eqn:eqn14}, that the deprojected (i.e. intrinsic) 3D mass distribution is subject to the mass-sheet transformation. The impact on the radial 3D mass distribution is highlighted in Fig. \ref{fig:fig2}, where we scale $\lint$ (and hence \dtint\ and $\kappa_{\rm mst,int}$) with respect to the mock input values. Differences in the total enclosed mass within \jwst's FOV amount to roughly 5\% for $\lint = 0.9$. Moreover, the mass-sheet transformation translates not only to differences in the total enclosed mass, but also changes the slope of the mass profile to which spatially resolved kinematics are sensitive \citep{2020MNRAS.494.4819C}.

%\comment{Sherry}{I think the equations below would be confusing for reader because we do not directly model $D_{\Delta t, {\rm composite}}$ in the L+D case, since we model $D_{\Delta t, {\rm int}}$.  When we wrote in our notes the expression using $D_{\Delta t, {\rm composite}}$, it was to see how various quantities scale with each other.  Also, it's not correct to include (1-$\kappa_{\rm ext}$) term here, without also transforming $\kappa_{\rm mst,int}$ to include another MST with $\kappa_{\rm ext}$.  I therefore recommend using $D_{\Delta t, {\rm int}}$ here and moving the (1-$\kappa_{\rm ext}$) to a separate discussion on the effect of external MST.}
%
%Incorporating further the effect of an {\it external} mass sheet \kext, we obtain 
%\begin{equation}
%\dt = \frac{\dtint}{(1-\kext)},
%\label{eqn:eqn17}
%\end{equation}
%whereas \dd\ is fully invariant under an \textit{external} MST \citep{2015JCAP...11..033J}, i.e. $\dd = \ddint$.

% SHS:
%the angular diameter distance to the lens is invariant under the MST \citep{2015JCAP...11..033J}, and hence
%\begin{equation}
%{D_{\rm d, \lambda}} = D_{\rm d}^{\rm model}
%\label{eqn:eqn4}
%\end{equation}
%Here, the "model" distances correspond to the inferred values of the composite model without any internal mass sheet contribution.
\subsection{Models}
\label{sec:models}

We construct joint strong lensing and stellar dynamical models as introduced in Y20. In brief, these models rely on a pixelated source fitting algorithm \citep{2006MNRAS.371..983S,2010ApJ...711..201S,2013ApJ...766...70S} and the solution of the Jeans equations in axisymmetry \citep[][JAMPY]{2008MNRAS.390...71C}. The model parameter space thus includes those listed in Table \ref{table:table1} for the \texttt{COMPOSITE} mass model and \texttt{SPEMD}, including the time-delay distance \dtint, the lens distance \ddint, the orbital anisotropy parameter $\beta_z$, inclination $i$ and absolute strength/amplitude of the internal mass-sheet parameter \lint. In line with Y20, we do not include the satellite in our fits to both the lensing and kinematic data. Given that the contribution of the satellite to the SMD is $\le$ 1\% at the lens centre, any uncertainties due to the missing satellite are subdominant with respect to the other sources of uncertainties in the models. Moreover, we do not constrain the parameter range by physical arguments other than the fact that the mass-sheet transformation is not supposed to yield negative convergences within a FOV that is more than twice as large as the core radius, which penalises models with $\lint \gtrapprox 1.05$\footnote{Our FOV within which we carry out the MGE has to be large enough to avoid a premature exponential fall off of the Gaussians, which can adversely impact the predicted \vlos. We find that a FOV $\ge 2\times \theta_{\rm c}$ is sufficiently large to adequately describe the profile of the cored density distribution. Within this FOV, however, the convergence profile is not allowed to be negative, since this would yield negative SMDs. This translates into an upper bound for our constraints on \lint\ of $\approx 1.05$.} (see Sec. \ref{sec:results}). Besides that, the priors are identical to those adopted in Y20, with \lint\ being flat within the aforementioned prior range of $[0.5, 1.5]$. Note that our kinematic data has been mocked up without any contribution from a mass sheet (i.e. $\lint = 1.0$, as described in Sec.~\ref{sec:data}).
Moreover, we do not assume a ratio of $D_{\rm d}/D_{\rm ds}$. That is, our inferred values for \dtint\ and \ddint\ are based solely on the data and are not constrained by assuming a cosmological world model.

\begin{table*}
	\caption{Model parameters and priors for our joint strong lensing \& dynamical models, including the cosmological distances, the \texttt{SPEMD}, the \texttt{COMPOSITE} mass distribution, the \texttt{PIEMD} (parameterising the mass sheet) and the dynamical variables, consisting of models with a constant ($\beta_{z}(r) = const.$) and spatially varying ($\rm{log_{10}}\,(\beta_{z}(r)) = \alpha + \beta \times\ \rm{log_{10}}\,(r)$) anisotropy profile. The mock IFU data set is based on the best-fitting \texttt{COMPOSITE} lensing-only model with a source resolution of $64\times 64$ pixels and arbitrary values for the dynamical parameters. The mock cosmological distances are based on the best-fitting lensing-only model for \dtint\ and assuming \h\ $ = 82.5$ \kmsM, $\Omega_{\mathrm{m}} = 0.27$, $\Omega_{\mathrm{\Lambda}} = 0.73$, $z_{\mathrm{d}} = 0.295$ and $z_{\mathrm{s}} = 0.654$ for determining \ddint.} 
	\begin{center}
	\centerline{
	\begin{tabular}{ c  c  c  c  c }
		\hline
		\hline
		Description & Parameters & Mock input values & Prior type & Prior range\\
		\hline
		\textbf{Distances} & & & & \\
		Model time-delay distance [Mpc] & \dtint\ & 1823.42 & Flat & [1000, 3000]\\
		Model lens distance [Mpc] & \ddint\ & 775.00 & Flat & [600, 1000]\\
		\hline
		\textbf{SPEMD} & & & & \\
		Flattening & $q$ & & Flat & [0.2, 1.0]\\
		Einstein radius [arcsec] & $\theta_{\mathrm{E}}$ & & Flat & [0.01, 2.0] \\
		Power law slope & $\gamma_{\rm pl}$ & & Flat & [0.2, 0.8]\\
		External shear strength & $\gamma_{\mathrm{{ext}}}$ & & Flat & [0.0, 0.2]\\
		External shear position angle [\textdegree] & $\phi_{\mathrm{ext}}$ & & Flat & [0.0, 360.0]\\
		\hline
		\textbf{COMPOSITE} & & & & \\
		Stellar M/L & \MLstar & 2.09& Flat & [0.5, 2.5]\\
		NFW flattening & $q$ & 0.73 & Flat & [0.2, 1.0]\\
		NFW Einstein radius [arcsec] & $\theta_{\mathrm{E}}$ & 0.20 & Flat & [0.01, 2.0] \\
		NFW scale radius [arcsec] & $r_{\mathrm{s}}$ & 22.53 & Gaussian & [18.6, 2.6]\\
		External shear strength & $\gamma_{\,\mathrm{ext}}$ & 0.08 & Flat & [0.0, 0.2]\\
		External shear position angle [\textdegree] & $\phi_{\mathrm{ext}}$ & 1.42 & Flat & [0.0, 360.0]\\
		\hline
		\textbf{PIEMD} & & & & \\
		Einstein radius [arcsec] & \lint & 1.00 & Flat & [0.5,1.5] \\
		\hline
		\textbf{Dynamics} & & \\
		Anisotropy & $\beta_{z}(r) = const.$ & 0.15 & Flat & [$-0.3$, 0.3]\\
		Inclination [\textdegree] & $i$ & 84.26 & Flat & [80.0, 90.0]\\
		\hline
		\textbf{Dynamics} & & \\
		Anisotropy & $\rm{log_{10}}(\beta_{z}(r)) = \alpha + \beta \times\ \rm{log_{10}}(r)$ &  &  & \\
		Pivot & $\alpha$ &  & Flat & [$-0.3$, 0.3]\\
		Slope & $\beta$ &  & Flat & [$-0.5$, 0.5]\\
		Inclination [\textdegree] & $i$ &  & Flat & [80.0, 90.0]\\
	\end{tabular}
	}
%	\vspace{2ex}
	\label{table:table1}
	\end{center}
	Note: the stellar M/L is the constant multiplied to the observed light intensity distribution in the mock imaging data to obtain the dimensionless surface mass density $\kappa$.  For a real system with real observations, this value can then be converted to physical units such as \Msun\ and \Lsun.
\end{table*}

We explore the posterior probability density function (PDF) by means of \textsc{emcee} \citep{2013PASP..125..306F}. We investigate different modelling runs by adopting different source grid resolutions, which (for a given mass model parameterisation and in the absence of a MSD) constitutes the biggest source of uncertainty on \h\ \citep[][Y20]{2013ApJ...766...70S}. The grid resolutions span a range of 56 $\times$ 56 to 68 $\times$ 68 pixels in the source plane, with source pixel sizes roughly of the order of 0.05$\pm$0.01 \arcsec/pixel. The grid size is sufficiently large to allow for a proper reconstruction of the extended source, without artificially constraining the parameter space in \lint, as will be shown for the lensing-only models in Sec. \ref{sec:results:distances}.  In a last step, we then combine the constraints from different source resolutions by assigning equal weights to the chains.% Finally, we employ the Bayesian Information Criterion \citep[][BIC]{schwarz1978}, where the kinematic likelihood is utilised to effectively rank our models (Y20).

In the case of the FIDUCIAL data set, we sample over an additional velocity fractional shift parameter $f_{\rm v}$ in the model to account for the systematic uncertainties. For a given value of $f_{\rm v}$, the mock kinematic data $\sqrt{\vlosldata}$ in Eq.\ref{eqn:eqn1} is scaled by $(1-f_{\rm v}) \sqrt{\vlosldata}$.  We consider a uniform distribution of $f_{\rm v} \in [-0.02,0.02]$, i.e., we assume the mock kinematic data in all kinematic bins have equal likelihood in being systematically shifted by an amount within 2\%. Such systematic fractional offset in the measured kinematic data could arise from, e.g., stellar template mismatches. In practice, such as systematic shift in the kinematic data leads to a shift in $\ddint$, leaving other model parameters unchanged. To obtain the cosmological distance measurements in the FIDUCIAL data set, we then marginalise over the parameter $f_{\rm v}$ from $[-0.02,0.02]$.

%============================= Section 4 =============================
\section{Results and Discussions}
\label{sec:results}
%=====================================================================

\subsection{Distance measurements without cosmological model assumptions}
\label{sec:results:distances}

\begin{figure*}
\begin{center}
\includegraphics[width=0.9\linewidth]{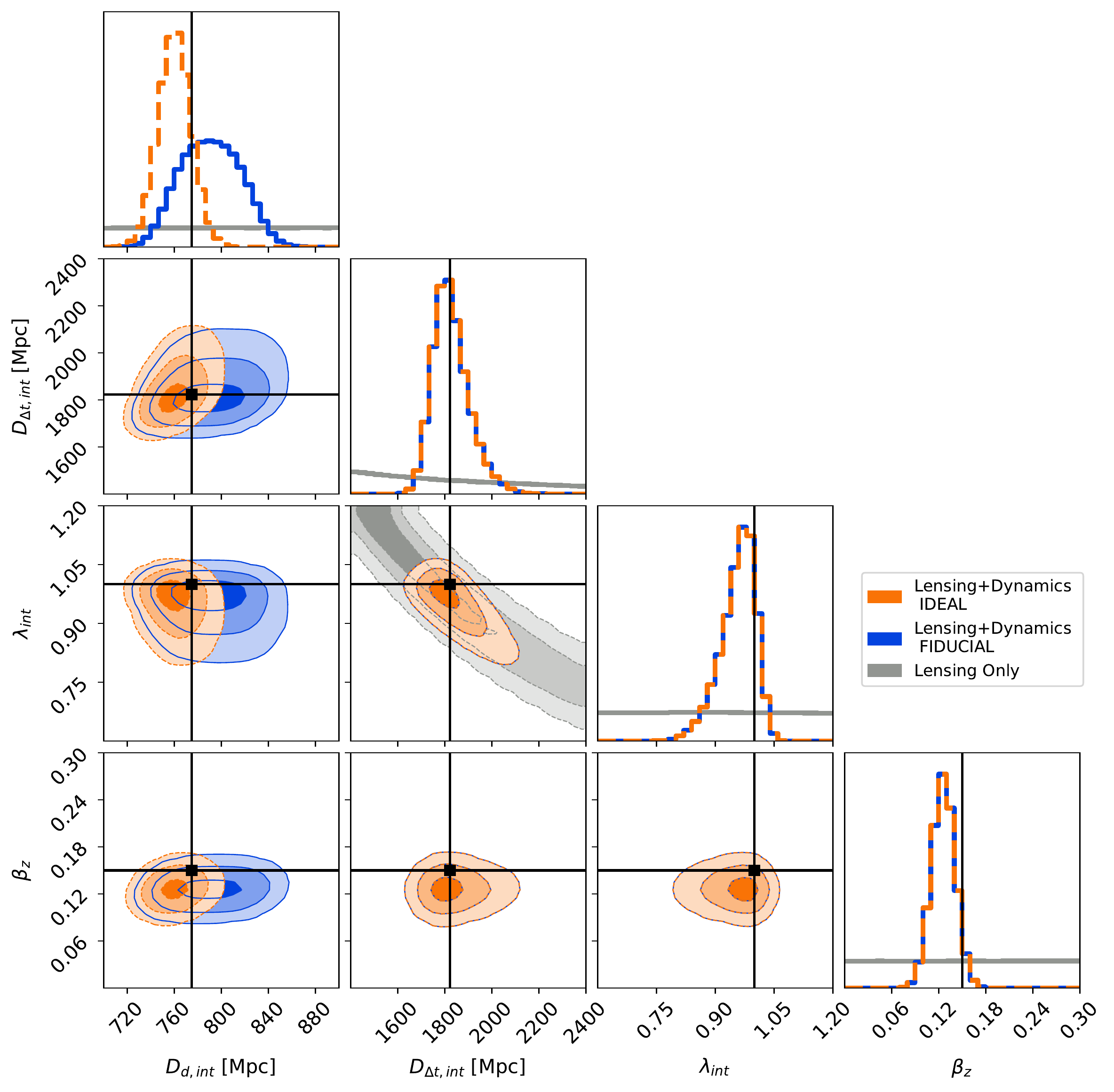}
\caption{Measurements from our joint strong lensing and stellar dynamical models, without any kinematic data (grey) and with mock IFU stellar kinematics (orange and blue). The corner plot shows the most relevant parameters in the fit, consisting of the lens distance \ddint, time-delay distance \dtint, internal mass-sheet parameter \lint\ and orbital anisotropy $\beta_z$. The contours highlight the 1-, 2- and 3$\sigma$ confidence regions. The black points (lines) depict the mock input values. Without any kinematics, \dtint\ and \lint\ are highly degenerate while \ddint\ and $\beta_{z}$ are unconstrained. The posterior PDFs for \ddint\ and $\beta_z$ are virtually flat in this case, which is why we only show the 2D contours in the \dtint-\lint\ space. IFU data will considerably tighten the constraints. In the IDEAL case, the mock input values are recovered within 2$\sigma$. In the FIDUCIAL case, the mock input values are also recovered  within the 2$\sigma$ bands, but with larger uncertainties for the lens distance ($\delta \ddint/\ddint \approx 4\%$). These measurements are independent of cosmological model assumptions.}
%The mean of the distribution is slightly offset from the actual input value, as our mock source resolution pick is an outlier in the combined distribution across all source resolutions.
\label{fig:fig3}
\end{center}
\end{figure*}

The probability distributions for the various data sets are displayed in Fig. \ref{fig:fig3}. As expected, the introduction of a constant \textit{internal} mass sheet is degenerate with the mass model. By scaling the mass model parameters accordingly, the lensing data (consisting of high-resolution imaging and time delays) can easily be recovered, while allowing for a large variation in the contribution of \lint. Consequently, we observe a posterior probability distribution function (PDF) for \lint\ and \dtint\ (grey) which is essentially flat within the prior bounds. The PDF for the lens distance \ddint\ and the anisotropy parameter $\beta_{z}$ is also flat, since these are anchored by the kinematic data only. It is worth noting here, that the effect depicted in Fig. \ref{fig:fig3} for the lensing-only data is neither new nor unexpected, but simply a manifestation of the MSD. The degeneracy is well captured by our models and this exercise also serves the purpose of illustrating the differences in the constraining power of a mass sheet with and without high-spatially resolved stellar kinematic data.

The corresponding probability distributions for our joint lensing and dynamical models are also shown in Fig \ref{fig:fig3}, where the IDEAL (orange) and FIDUCIAL (blue) mock IFU observations were employed. In contrast to the highly degenerate lensing-only run, an internal mass-sheet component is constrained when IFU stellar kinematics are available. We observe tight constraints for any contribution from an internal mass sheet with a core radius of 20\arcsec\ of $\lint = 0.97^{+0.03}_{-0.05}$. The sensitivity of the 2D kinematics to changes in the radial mass distribution translates to tight measurements on the distances, too. We measure \ddint\ and \dtint\ with a precision of $\sim$ 1.6\% and $\sim$ 3.7\% respectively (see Table \ref{table:table2}) in the IDEAL case. Yet, the constraint for the lens distance degrades significantly when the FIDUCIAL data is employed instead, where \ddint\ is only measured with a $\sim$ 4\% precision. In both cases, however, the mock input values are generally recovered within the 1-2$\sigma$ uncertainty bands. The results can be explained as follows; the statistical uncertainty in the mock kinematic data ($\delta v_{\rm{stat}}$) sets the overall precision for measuring \lint\ and \dtint, whereas the systematic offset ($\delta v_{\rm{sys}}$) dictates the best precision to be achieved for \ddint. Hence, with the same statistical uncertainties of 2.5\% in both the IDEAL and FIDUCIAL data, the final modelling uncertainties for \lint\ and \dtint\ are identical (see Fig. \ref{fig:fig3}), while the uncertainties on \ddint\ in the FIDUCIAL case can readily be obtained from the IDEAL PDF, by utilising the analytic relationship in Eq. \ref{eqn:eqn20}\footnote{In detail, the final uncertainties for \ddint\ in the FIDUCIAL case are obtained by marginalising over models with various realisations of the systematic offset $f_{\rm v}$. In practice, we create 9 realisations corresponding to the values of $f_{\rm v}=\{-0.02, -0.015, -0.01, -0.005, 0., 0.005, 0.01, 0.015, 0.02\}$, and combine the resulting chains from these realisations with equal weights.}.
%That is, we create uniform realisations of the mock FIDUCIAL data with a bias of $-2\pm2\%$, where the individual chains are combined equally to 
This yields the final PDF in Fig. \ref{fig:fig3}.

While we recover the key parameters of our mass models within the 1-2$\sigma$ regions as shown in Fig.~\ref{fig:fig3}, we note here some of the effects that affect the shape of the posterior PDF of the model parameters. First of all, the amount of noise in the mock data sets the size of the posterior PDF -- the higher the amount of noise, the broader the PDF (as further illustrated in Appendix \ref{sec:appendixa}).  Secondly, the posterior distribution of $\lint$ is highly asymmetric due to the imposed physical condition of positive mass density.  Finally, models with different source pixelisations lead to slightly different optimisations, resulting in shifts of the PDF from individual source pixelisations relative to each other \cite[][Y20]{2013ApJ...766...70S}. The final PDF is obtained by combining the individual PDFs for each source resolution, giving equal weighting to all due to indistinguishable kinematic goodnes of fit values ($\chi_{\rm D}$), in order to account for the model uncertainties due to source pixelisation. This is in contrast to Y20, where the Bayesian Information Criterion \citep[BIC;][]{schwarz1978} was utilised effectively to downweight individual source resolutions that are too far away from the mock input values, but no longer applicable in the presence of a MST to further constrain this source of uncertainty in the models.

For the orbital anisotropy, we picked a mock input value of $\beta_{z} = 0.15$. Our choice of mild radial anisotropy was motivated by the findings from extensive modelling of massive elliptical galaxies in the local universe \citep{2007MNRAS.379..418C,2013MNRAS.432.1709C}. Regardless of this particular choice, Fig. \ref{fig:fig3} is a powerful illustration of how time delays and high-spatially resolved kinematic data will enable us to lift the MSD and mass-anisotropy degeneracy \citep{1993ApJ...407..525V,1993MNRAS.265..213G} simultaneously, providing tight constraints also on the orbital anisotropy parameter, which is otherwise difficult unless higher order moments of the line-of-sight velocity distribution are available. 

\begin{figure*}
\begin{center}
\includegraphics[width=0.9\linewidth]{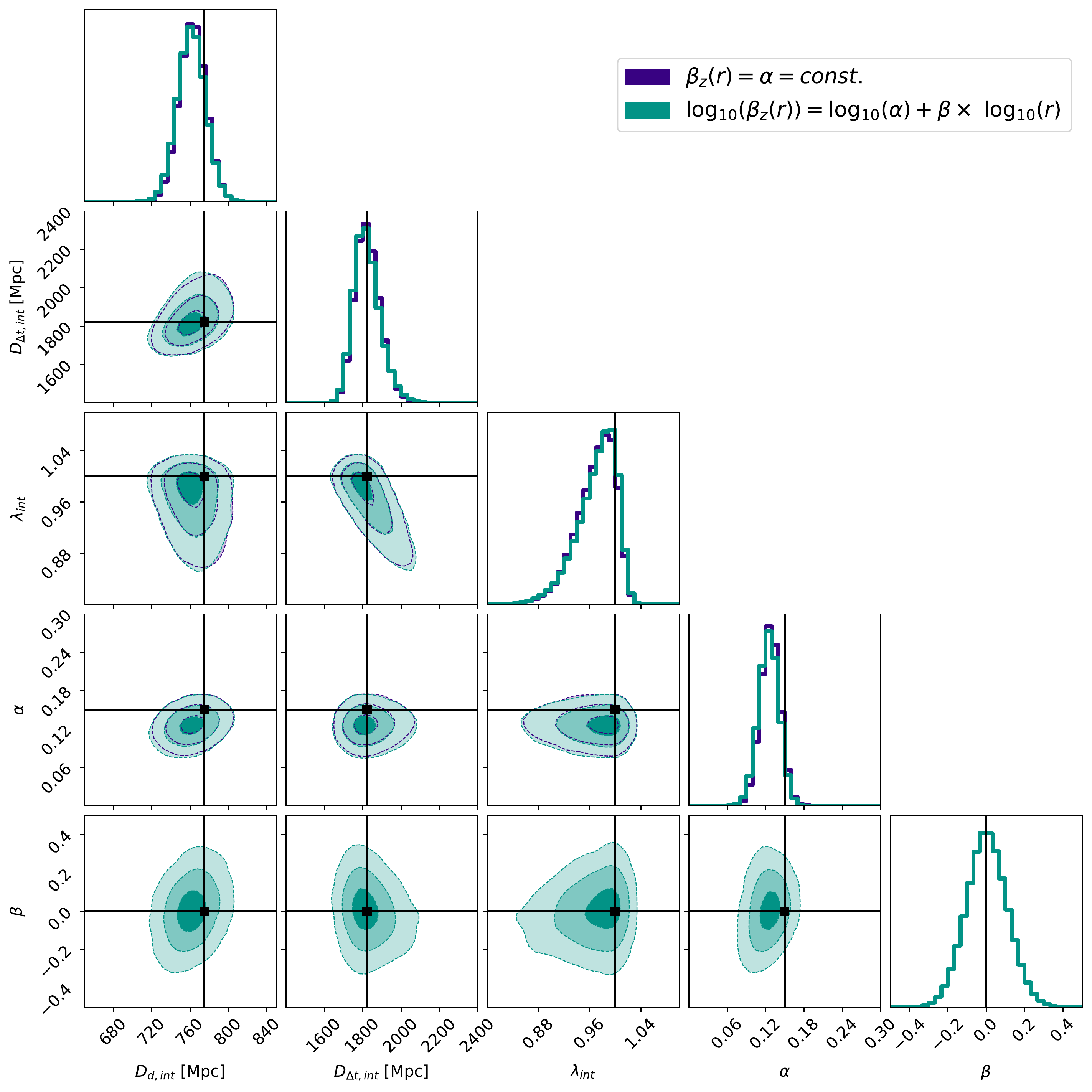}
\caption{Measurements from our joint strong lensing and stellar dynamical models with mock IFU stellar kinematics (IDEAL), but for two different parameterisations of the orbital anisotropy profile $\beta_{z}$. The corner plot shows the relevant parameters in the fit, consisting of the lens distance \ddint, time-delay distance \dtint\ and internal mass-sheet parameter \lint. The mock data is mildly anisotropic (i.e., $\beta_{z}(r) = 0.15$ is non-zero and independent of radius) and fitted with both a radially constant (indigo) and radially varying anisotropy profile (teal). The contours highlight the 1-, 2- and 3$\sigma$ confidence regions, with the black points (lines) depicting the mock input values. The radially constant mock input anisotropy can be recovered from models with a radially varying anisotropy profile.}
%The mean of the distribution is slightly offset from the actual input value, as our mock source resolution pick is an outlier in the combined distribution across all source resolutions.
\label{fig:fig4}
\end{center}
\end{figure*}

The anisotropy in our mock data and models was, so far, assumed to be constant as a function radius.  We also explored the scenario of allowing for spatially varying anisotropy profiles to assess its impact on the inference of the cosmological distances in the fit. We present the results in Fig. \ref{fig:fig4}, where we compare the constraints for the two different parameterisations of the orbital anisotropy profile. Models which adopt a constant value for the anisotropy fit the data exceptionally well and recover the mock input values within the 1$\sigma$ confidence levels (indigo). This is identical to Fig. \ref{fig:fig3}, but shown here for a single source resolution of $64 \times\ 64$ pixels. Models with a radially varying anisotropy profile have been implemented, where the anisotropy profile is parameterised according to:
\begin{equation}
\rm{log_{10}}\,(\beta_{z}(r)) = \rm{log_{10}}\,(\alpha) + \beta \times \rm{log_{10}}\,(r),
\label{eqn:eqn21}
\end{equation}
with $\alpha \in [-0.3,0.3]$ and $|\,\beta\,| \in [0,0.5]$.
This parameterisation provides enough freedom to allow for a wide range of anisotropy profiles, while ensuring a smooth transition from the central to the outermost bins. It is also tailored to account for the coarser sampling in the remote regions. Models with a radially varying anisotropy profile are capable of recovering the mock input value within 1$\sigma$ (teal), with $\alpha = 0.13\pm0.02$ and $\beta = 0.0\pm0.1$. Despite the increased freedom in the anisotropy profile, we do not observe inflated errors for the cosmological distances in the fit, showing that the orbit distribution has limited power for compensating the effects of the distances and mass-sheet, as illustrated in Fig. \ref{fig:fig2}.

\begin{table*}
	\caption{Parameter constraints from our lensing-only and joint strong lensing \&\ stellar dynamical models. The IFU stellar kinematics have been mocked with (FIDUCIAL) and without (IDEAL) systematic errors included and a statistical uncertainty of $\delta v_{\rm{stat}}$ = 2.5\% per bin. The last column represents the inferred \h\ values for the various models, under the assumption of flat \lcdm. The mock input value is 82.5 \kmsM. We quote the 50th and 16th/84th percentiles of the distribution.}
	\begin{center}
	\centerline{
	\begin{tabular}{ c | c  c  c c }
		\hline
		Model & \ddint\ [Mpc] & \dtint\ [Mpc] & \lint & \h$\,$[\kmsM] (flat \lcdm) \\
		\hline
		Lensing & - & 1749$^{+897}_{-891}$ & 1.00$^{+0.50}_{-0.50}$ & 82.7$^{+24}_{-22}$ \\
		\\
		Lensing \& Dynamics & 761$^{+12}_{-12}$ & 1816$^{+81}_{-53}$ & 0.97$^{+0.03}_{-0.05}$ & 84.1$^{+2.8}_{-2.6}$ \\
		IDEAL with $\delta v_{\rm stat} = 2.5\%$\\\\
		\\
		Lensing \& Dynamics & 791$^{+26}_{-26}$ & 1816$^{+81}_{-53}$ & 0.97$^{+0.03}_{-0.05}$ & 82.2$^{+3.1}_{-3.1}$ \\
		FIDUCIAL with $\delta v_{\rm stat} = 2.5\%$\\\\
		\\
        % \hline
	\end{tabular}}
\end{center}
%	}
%	\vspace{2ex}
	\label{table:table2}
%	\end{center}
\end{table*}

\begin{figure*}
\begin{center}
\includegraphics[width=0.99\linewidth]{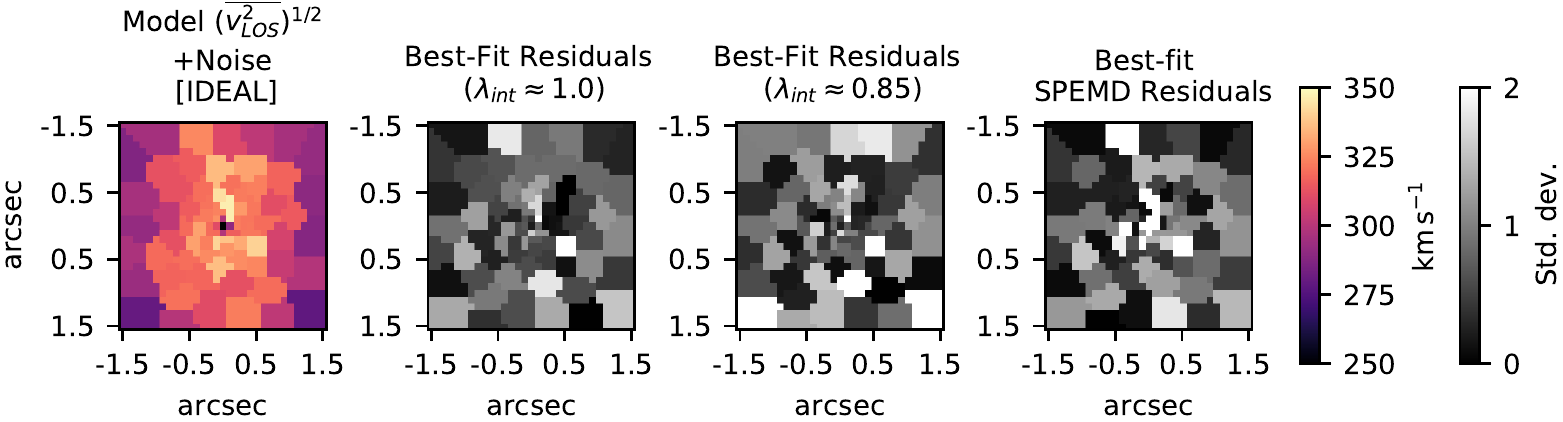}
\caption{First: IDEAL mock IFU \vlos\ data of \rxj, at \jwst\ NIRSpec resolution. Following the procedure in Y20, the data has been mocked up by means of the best-fitting lensing-only model and a set of random dynamical parameters. Second: normalised residuals (in absolute values) of the best-fitting model without a mass-sheet component ($\lint \approx 1.0$), showing that the mock data is remarkably well recovered mostly within the 1-$\sigma$ uncertainties. Third: normalised residuals (in absolute values) of the best-fitting model with a (fixed and significant) mass-sheet component ($\lint \approx 0.85$), which is at the edge of the 3-$\sigma$ uncertainty bands. Fourth: same as the two panels before, but for the best-fitting model with a \texttt{SPEMD}. Various mock data points are not recovered even within the 2-3$\sigma$ uncertainties. The $\chi^{2}$ difference between this model and the best-fitting \texttt{COMPOSITE} model without a sheet (second panel) is $\Delta\chi^{2} \approx 155$.}
\label{fig:fig5}
\end{center}
\end{figure*}

We further highlight the difference in the 2D kinematics between the IDEAL mock data, the best-fitting model without a mass-sheet ($\lint \approx 1.0$) and the best-fitting model with a (fixed) mass-sheet of $\lint \approx 0.85$ in Fig. \ref{fig:fig5}. The mock data is exceptionally well fitted without a mass sheet and generally recovered within the 1-2$\sigma$ errors. A value of $\lint \le 0.9$ is deemed necessary to align the \holi\ \h\ constraints of \rxj\ with those from the cosmic microwave background (B20), but such significant deviations from lens galaxies without a sheet can be ruled out at the $\sim 3 \sigma$ level, based on our IDEAL mock kinematics. The difference between the overall best-fitting model without a mass sheet (i.e. $\lint = 1.0$) and the best-fitting model with a significant mass-sheet contribution of $\lint \approx 0.85$ is $\Delta\chi^{2} = 6$. Given the same priors for the two models above, the ratio of the posterior probabilities is determined by the Bayes factor, which we approximate by means of the BIC difference (Y20), and which yields a posterior odd of 0.05 for the model with a mass sheet against the model without a mass sheet.  %In comparison, the FIDUCIAL IFU data still helps in ruling out a significant mass-sheet component. But, the differences between the data and best-fitting model with a massive sheet ($\lint \ge 0.9$) are minuscule in this case, given the larger error bars in the FIDUCIAL data set.

We also constructed models with a softened power-law elliptical mass distribution \citep[\texttt{SPEMD};][]{1998ApJ...502..531B} and, again, added a \texttt{PIEMD} profile to mimic the contribution of an internal mass-sheet. In agreement with our earlier studies (Y20), the \texttt{SPEMD} provides a considerably worse fit to the mock kinematics. The best-fitting \texttt{SPEMD} $\chi_{\rm D}^{2} \approx 235$, as opposed to $\chi_{\rm D}^{2} \approx 78$ for the \texttt{COMPOSITE} mass model. Whereas the BIC differences are insufficient to discriminate between the various source resolutions of a given mass model, the BIC differences are sufficient to rule out mass density parameterisations that deviate strongly from the \textit{input} model, even in light of the increased freedom provided by the MSD. Therefore, including the SPEMD models in our final chains leaves the constraints for the various parameters in the fit unchanged, given that these are assigned zero weighting.

The final parameter constraints for \ddint, \dtint\ and \lint\ are summarised in Table \ref{table:table2}. In concordance with Y20, we observe tight measurements for \ddint\ and \dtint\ when IFU data is available. The uncertainties on \ddint\ are dominated by those of the kinematic data, yet significantly better than expected from e.g. a single aperture averaged velocity dispersion \citep{2020arXiv201106002C,2019Sci...365.1134J}, given the roughly 80 (almost) independent measurements of the LOSVD across the entire FOV. The precision of \dtint\ also greatly benefits from high-spatially resolved kinematics (Y20), with $|\delta\dtint/\dtint| \sim |\delta\lint/\lint|$, but is highly sensitive to the statistical errors in the measurement of the LOSVD. For illustration purposes, we also present an even more conservative scenario in Appendix \ref{sec:appendixa}, in which we increase $\delta v_{\rm{stat}}$ to 3.5\% instead, which further highlights the impact of $\delta v_{\rm{stat}}$ in constraining \lint\ and \dtint.

\subsection{Cosmographic forecast in flat \lcdm}
\label{sec:results:forecast}

Unlike earlier findings without an \textit{internal} MST, where much of the constraining power in measuring \h\ came from \dtint\ only (with the small $\delta\dtint$ propagating almost linearly into the error budget on \h), \ddint\ is now expected to add substantial constraining power to \dtint\ in measuring \h\ \citep{2016JCAP...04..031J,2020arXiv201106002C}. 

To this end, we utilise the joint distance measurements of \ddint\ and \dtint\ to obtain cosmological forecasts in flat \lcdm. We first fit the marginalised \dtint-\ddint\ distribution (from all chains across various source resolutions) with a kernel density estimator and account for mass structures along the lens LOS with an external convergence (\kext) distribution, following the number counts approach from numerical simulations in \cite{2013ApJ...766...70S,2014ApJ...788L..35S}.
%SHS: %The posterior probability distribution for \dd\ and \dt\ is then related to a flat $\Lambda$CDM model through importance sampling, 
We draw random samples of \{\h, $\Omega_{\rm m}, \kext\}$ with uniform priors on \h\ of [50,120]\,\kms\ and on $\Omega_{\rm{m}}$ of [0.05, 0.5] and with the $P(\kext)$ distribution, compute the corresponding \dtint\ and \ddint\ following Eqs.~\ref{eqn:eqn7} and \ref{eqn:eqn8}, and importance sample given the $P(\ddint, \dtint)$ posterior. This yields a final value of \h\ $= 82.2^{+3.1}_{-3.1}$\;\kmsM\ (\h\ $= 84.1^{+2.8}_{-2.6}$\;\kmsM), i.e. a 3.2\% precision measurement (3.7\%) for the IDEAL (FIDUCIAL) case, recovering the input value of 82.5\,\kmsM. 

\begin{figure*}
\begin{center}
\includegraphics[width=0.49\linewidth]{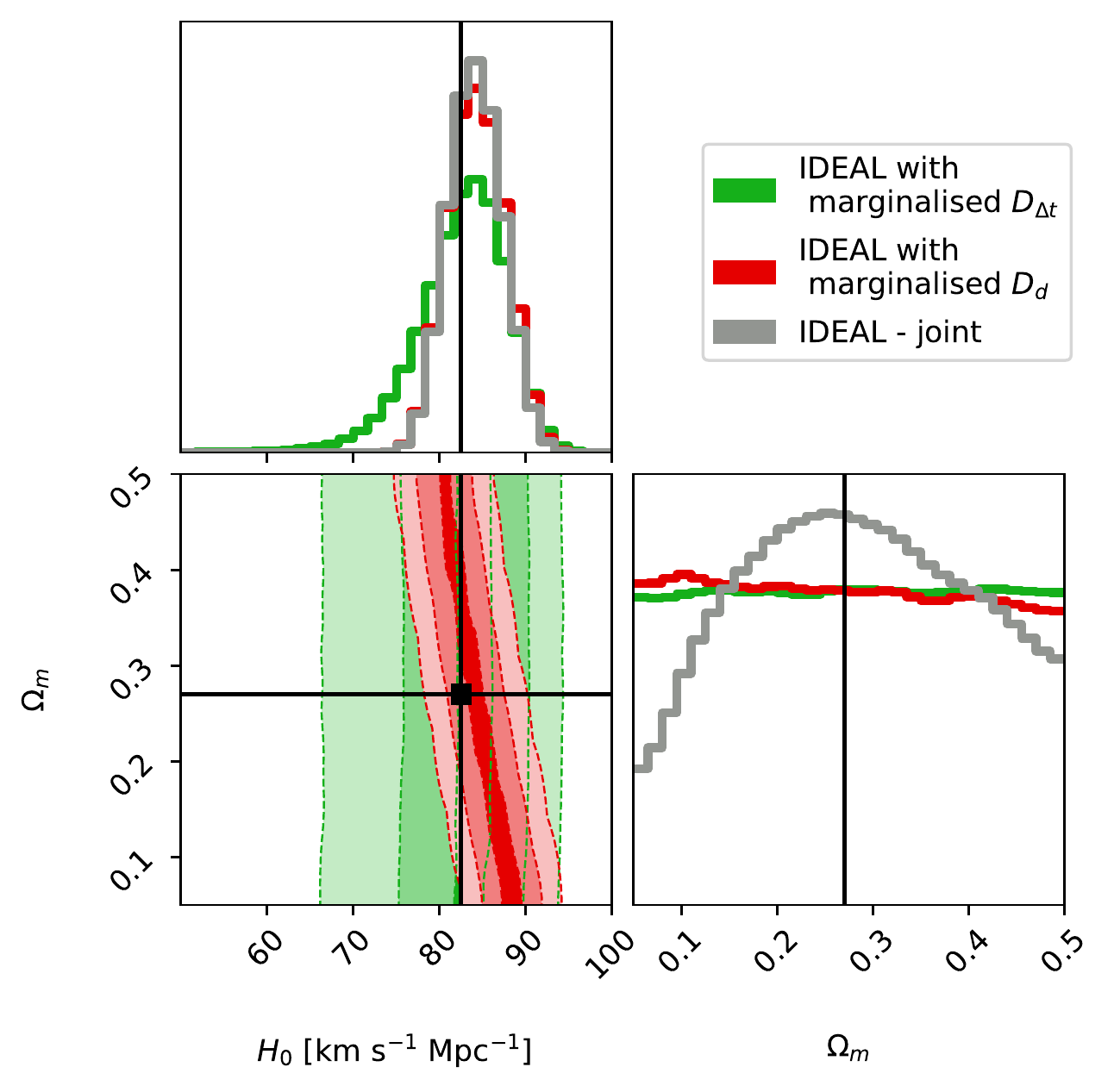}
\includegraphics[width=0.49\linewidth]{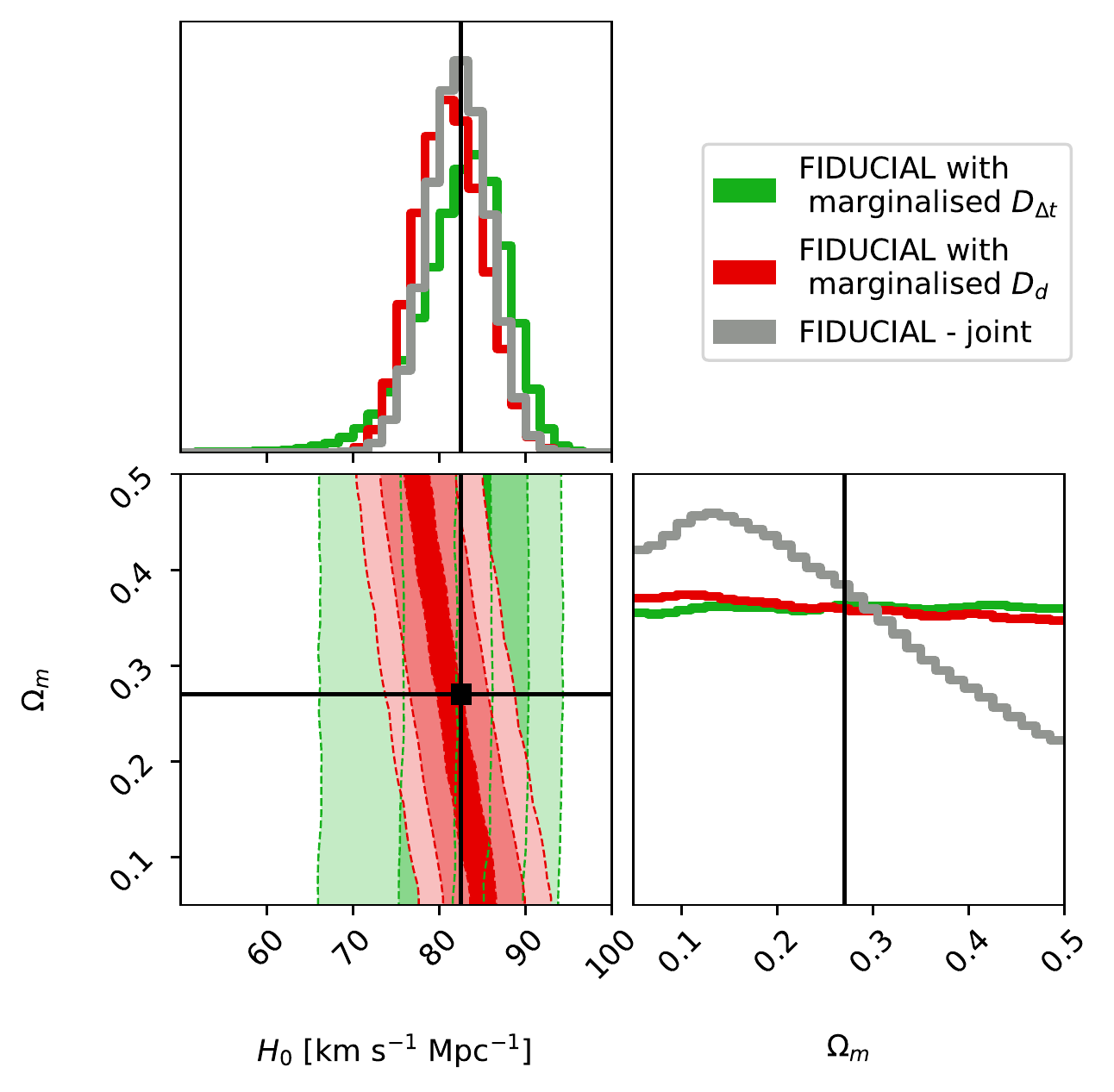}
\caption{\h\ and $\Omega_{\rm{m}}$ constraints from our models in flat \lcdm\ cosmology, for the IDEAL (left) and FIDUCIAL (data set) with statistical uncertainties of $\delta v_{\rm{stat}}$ = 2.5\%. We adopted uniform priors on \h\ of [50, 120]\,\kmsM and $\Omega_{\rm{m}}$ of [0.05, 0.5]. The green shaded regions show the 1, 2 and 3$\sigma$ confidence regions 
%for \dt\ marginalised models, i.e. effectively lensing-only models.
based on the constraint on $\dt$ only, i.e., $P(\dt)$ after marginalising over other mass model parameters including $\dd$.
The red shaded regions show the corresponding constraints %marginalised over \dd. 
based on $P(\dd)$, after marginalising over $\dt$. The grey histogram shows the constraints based on $P(\dd,\dt)$, where we have omitted showing the 2D contours for ease of viewing.
Different tilts in the \h-$\Omega_{\rm{m}}$ space, when marginalised over \dd\ and \dt, respectively, can break some of the degeneracies, leading to improved constraints also on $\Omega_{\rm{m}}$.}
%the dark matter density. 
%Note that given the smaller overlap of the \dd\ and \dt\ marginalised models in the FIDUCIAL data case yields to slightly tighter constraints on the joint \h\ inference.}
\label{fig:fig6}
\end{center}
\end{figure*}

Fig. \ref{fig:fig6} shows how the inclusion of 2D stellar kinematics of the lens not only helps in tightly constraining the lens distance, but also leads to improved constraints on cosmological parameters beyond \h, even for a single lens. This conclusion has already been drawn in Y20 and is now corroborated even in the presence of a MST, assuming that the stellar kinematics can be measured with high accuracy and precision. Otherwise, more systems will be needed to provide more meaningful constraints beyond \h.

\subsection{View of \dd\ and core size}
\label{sec:results:views}

%\akinedit{The results presented here rest upon the assumption of an extended, yet finite core size of $r_{\rm c} = 20\arcsec$. It is of great importance, however, 
It is of great importance to recapitulate our findings within a more comprehensive and broader context: despite employing a model that is maximally degenerate with \dtint, and hence with \h\ through the MST, we have shown that high-quality, high-spatially resolved stellar kinematics can break the \textit{internal} MSD and constrain \dtint. 
Furthermore, time-delays and (spatially resolved) kinematics also provide leverage for constraining \dd\ \citep{2009A&A...507L..49P} regardless of the MST \citep{2015JCAP...11..033J,2019Sci...365.1134J,2020arXiv201106002C}.  Previous analyses in TDCOSMO using two families of mass models with single-aperture velocity dispersion measurements and without MST yield joint constraints on \dt\ and \dd, but with substantially tighter constraints on \dt\ compared to \dd.  When allowing for a mass-sheet transformation, the precision on \dt\ degrades.  On the other hand, \dd\ is nearly insensitive to the internal MST (and completely insensitive to the external MST).  With high-quality spatially-resolved kinematics, the precision on \dd\ can become better than that of \dt\ (Sec.~\ref{sec:results:distances}, \ref{sec:results:forecast} and Y20).  Therefore, the \dt-centric view in TDSL can be complemented by the new \dd-centric view; namely, the combination of the two distances provide robust $H_0$ even in the presence of the MST.
%This \dt-centric view in TDSL can be complemented by the new \dd-centric view; namely that time-delays and (spatially resolved) kinematics also provide leverage for constraining \dd\ \citep{2009A&A...507L..49P} regardless of the MST \citep{2015JCAP...11..033J,2019Sci...365.1134J,2020arXiv201106002C}. %
%These two views are not independent but connected, and we have discussed the precision of \dd\ from \jwst-like stellar kinematics and its impact on the cosmological inference in Sec.~\ref{sec:results:distances}, \ref{sec:results:forecast} and Y20.

The results presented here rest upon the assumption of an extended, yet finite core size of $\theta_{\rm c} = 20\arcsec$, but it may be useful to emphasise again the role played by \dd\ in the two limiting cases of $\theta_{\rm c} \ll \theta_{\rm E}$ and $\theta_{\rm c} \gg \theta_{\rm E}$. First, when $\theta_{\rm c} \ll \theta_{\rm E}$, the approximation of a pure MST is no longer valid \citep{2020A&A...643A.165B} and lensing data are sensitive to transformations of the lens mass profile, particularly the $\lint$ values.  In this scenario, the precision of \dt\ is mainly limited by the precision of the lensing and time delay observations, whereas the precision of \dd\ mainly depends on the time-delays and kinematics. Both \dt\ and \dd\ can be measured almost equally well in this case (see also Y20), thus leading to tight constraints on the inferred cosmological parameters.
Second, when $\theta_{\rm c} \gg \theta_{\rm E}$, the kinematic data no longer constrains the internal mass sheet, as the corresponding density profile effectively vanishes (Appendix \ref{sec:appendixc}). Under these circumstances, no improvement in the measurement of \dt\ and \h\ can be expected, leaving it prone to the full degeneracy introduced by the MST. However, the kinematics still improve the constraint on \dd\ significantly on a single lens basis, which then becomes the dominant source of information in the final cosmological inference. While this was expected from previous studies on the subject \citep{2015JCAP...11..033J,2016JCAP...04..031J,2019Sci...365.1134J,2020arXiv201106002C}, our results demonstrate this explicitly and quantitatively using mock \jwst\ observations.

\subsection{Comparison to previous TDCOSMO studies}
\label{sec:results:comparison}

We compare our approach to previous analyses of TDCOSMO, particularly TDCOSMO I \citep{2020A&A...639A.101M}, IV \citep{2020A&A...643A.165B}, V \citep{2021A&A...649A..61B} and VI \citep{2020arXiv201106002C} that involve lens mass modelling.We start by summarising previous studies, and explain the similarities and differences.

In TDCOSMO I, \citet{2020A&A...639A.101M} investigated potential sources of systematic uncertainties connected to lens mass modelling by employing two families of physically motivated mass models, the power-law and composite mass models, without internal MST. They find the two families of models yield the same $H_0$ within 1\% uncertainty from the sample of lens systems that were previously modeled by the \holi, SHARP and STRIDES collaborations \citep{2013ApJ...766...70S,2014ApJ...788L..35S,2017MNRAS.465.4895W,2019MNRAS.484.4726B, 2019MNRAS.490.1743C,2020MNRAS.498.1440R,2020MNRAS.494.6072S}.  The comparison of these two families of model demonstrates that there is no evidence for significant residual systematic uncertainties associated with the lens mass modelling.

To further explore the MST as a source of uncertainty for measuring $H_0$, \citet{2020A&A...643A.165B} in TDCOSMO IV relaxed the assumptions on the lens mass profiles and allowed for an internal MST of the power-law model.  The MST parameter $\lint$ is then constrained through spherical Jeans modelling (with a radially varying anisotropy profile) and the stellar kinematic data of the lens galaxy, which are mostly single-aperture averaged stellar velocity dispersions. A subset of the SLACS lens galaxies is further combined with the 7 TDCOSMO lenses through a hierarchical analysis to constrain $\lint$ and infer $H_0$ in flat $\Lambda$CDM.  The uncertainties on $H_0$ are 8\% from TDCOSMO lenses only, or 5\% from TDCOSMO and SLACS lenses. \citet{2021A&A...649A..61B} in TDCOSMO V then estimated the precisions of $H_0$ measurements from current and future samples of time-delay lenses, with various levels of velocity dispersion measurements, ranging from single aperture-averaged velocity dispersions to IFU kinematic maps with 30 radially averaged velocity bins.  With 7 TDCOSMO lenses and future velocity dispersion measurements in 10 radial bins for each lens, \citet{2021A&A...649A..61B} anticipate a 3.3\% precision on $H_0$ in flat $\Lambda$CDM.  An expansion of the sample to 40 time-delay lenses and 200 non-time-delay lenses would improve the precision to 1.5\% and 1.2\%, respectively.

The analyses of TDCOSMO IV and V assumed the flat $\Lambda$CDM model, which reduces the range of plausible $\lint$ in the MST given the tight correlation between \dd\ and \dt.  \citet{2020arXiv201106002C} in TDCOSMO VI relaxed the assumption of flat $\Lambda$CDM by considering the measurements of \dt\ and \dd\ in generic cosmological models, while allowing for the MST.  By combining the lenses together with relative distance indicators such as Type Ia Supernovae and Baryon Acoustic Oscillations, \citet{2020arXiv201106002C} showed that $\lint$ and $\dt$ can be constrained in generic cosmological models. In this work, spherical Jeans modelling and aperture-averaged kinematic measurements are used.

Our current paper is similar to TDCOSMO IV, V, VI in that we have also considered a parametrised profile with an internal mass-sheet transformation characterised by $\lint$.  While TDCOSMOS IV, V, VI used only the power-law profile with mass sheets, we have considered both families of mass models (composite and power-law profiles that were used in TDCOSMO I) with additional mass sheets.  We find that spatially resolved kinematics not only allow us to constrain the MST parameter $\lint$, but also to distinguish between the families of mass models.  Our set up is similar to TDCOSMO VI in that we aim to measure the distances $\dt$ and $\dd$ without assuming a background cosmological model, in contrast to TDCOSMO IV and V that assumed flat $\Lambda$CDM.

The key difference and novelty of our current work are the use of a more flexible, self-consistent lensing and dynamical mass model.  Instead of spherical Jeans modelling for the kinematics that were used in previous TDCOSMO studies, we have employed an axisymmetric mass distribution (in 3D) for both the lensing and kinematic analysis.  Furthermore, we have considered 2D spatially resolved kinematic data with angular structure, which help to constrain both $\lint$ and $\dt$, even when allowing for radially varying anisotropy profiles. With our forecast of JWST-like spatially resolved kinematics of $\sim$80 bins for \rxj, we find that we can constrain $H_0$ with this single lens with better than 4\% precision. This is based on the assumption of a uniformly distributed systematic uncertainty of 2\% with $f_{\rm v}$ varying between [-0.02,0.02], although we show in Appendix \ref{sec:appendixb} that a Gaussian distributed systematic uncertainty of 2\% increases the $H_0$ uncertainty only slightly from 3.8\% to 4.1\% for our mock kinematic data with $\delta v_{\rm{stat}}$ = 2.5\%.  In comparison to TDCOSMO V that adopted a 3\% systematic uncertainty that is larger than our 2\%, our forecast of $\leq 4\%$ uncertainty on $H_0$ constrained from a single lens is more precise, as expected.  Once we account for the difference in the assumed kinematic data quality, the two studies (this and TDCOSMO V) should provide consistent results.  A direct comparison of the two dynamical modeling approaches (axisymmetric Jeans in this paper and spherical Jeans in TDCOSMO V) would require modeling the same kinematic, lensing and time delay data set, which is not the focus of our current paper.

%This is more precise per lens compared to the forecast in TDCOSMO V. However, once we account for the difference in the assumed kinematic data quality, the two studies (this and TDCOSMO V) should provide consistent results.}

One other key difference to TDCOSMO IV, V and VI is that with the anticipated high-quality JWST kinematic data, we can constrain the parameters for the mass sheet ($\lint$) and for the anisotropy of the stellar orbits from the JWST kinematic data, without having to rely on external samples of non-time-delay lenses (in TDCOSMO IV and V) and/or other relative distance indicators (in TDCOSMO VI). We therefore see the individual analyses of lenses with high-quality data to be complementary to the hierarchical analyses with external samples -- the combination of both approaches in future studies would give us the maximal amount of cosmological information especially in measuring $H_0$.

%============================= Section 5 =============================
\section{Summary}
\label{sec:summary}
%=====================================================================

In order to assess the constraining power on a mass-sheet component in TDSL studies, we have performed a suite of simulations. For this purpose, we used the prominent lens system \rxj\ as a test-bed for our studies. %By mocking up high-quality, high-spatially resolved IFU stellar kinematics (as is expected from the next generation of ground and space-based telescopes), we have remodelled the data, including the %archival
%\akinedit{mock} \hst\ imaging and long-baseline time-delay measurements. 
We mocked up high-quality, high-spatially resolved IFU stellar kinematics (as is expected from the next generation of ground and space-based telescopes) together with mock \hst\ imaging, and used long-baseline time-delay measurements. 
Following the prescription of a cored density distribution (B20), a variable \textit{internal} mass-sheet component was included in our models. Our findings can be summarised as follows:

\begin{enumerate}
\item
Without any kinematic data (i.e. for pure strong lens models), we observe the well known degeneracy between the mass model parameters and any internal mass sheet component \lint. The lensing data allow for a wide range of \lint\ values, which can adversely affect the time-delay distance measurement \dtint, according to Eq. \ref{eqn:eqn6}.
\item Future IFU stellar kinematics with the next generation of telescopes (such as \jwst\ and \elt) will be able to effectively break the \textit{internal} MSD, by placing tight constraints on the contribution of any internal mass-sheet component with a core radius $\lessapprox 20\arcsec$ (This finding is specific to \rxj, However, with $\theta_{\rm E}/\Reff \approx 0.9$, \rxj\ is not a particular outlier in the \holi\ sample \citep{2020A&A...639A.101M}.).
\item Using realistic mock IFU observations, provided by assuming reasonable on-source integration times with \jwst, we show that future measurements of \lint\ (and hence \dtint) could be as precise as 4\% {\it per lens}. These uncertainties propagate almost linearly into the error budget on \h, whereas the precise \ddint\ measurements ($\delta \ddint/\ddint \le 4\%$) allows further reduction in the \h\ uncertainties. This yields a high-precision measurement on \h\ of $\le 4\%$ from this single lens system, assuming flat \lcdm\ cosmology.
\item High-spatially resolved stellar kinematics of the lens can constrain cosmological parameters beyond \h. However, more systems are needed to provide more meaningful information on e.g. $\Omega_{m}$.
\item The modelling approach presented here is fully self-consistent and more flexible than literature studies where spherical symmetry is commonly employed for modelling the stellar kinematic data. The 2D angular structure in the kinematic data together with the lensing information allow us to recover the anisotropy parameters of the stellar orbits, even with radial variations in the stellar anisotropy parametrisation.
%Spatial variations in the stellar mass-to-light ratio and/or orbital anisotropy can also easily be adopted, without resorting to specific functional forms \citep{1979PAZh....5...77O,1985MNRAS.214P..25M,1985AJ.....90.1027M}. Although a more detailed study would be necessary to assess possible systematics related to e.g. the intrinsic shape of lens galaxies, we expect this implementation to be less susceptible to certain modelling assumptions. 
\item We do not rely and make no use of informative priors at the galaxy population level. This approach presents an alternative and complementary method to that presented in \cite{2020A&A...643A.165B}, to gain back much of the precision lost by accounting for the MSD. In addition, spatially-resolved kinematics can be used, in turn, to verify if external data sets can be combined with the TDCOSMO lenses from the galaxy population point of view.
\item Our final inference of the angular diameter distances \ddint\ and \dtint\ are independent of cosmological model assumptions.
\end{enumerate}

It is beyond the scope of this paper to speculate on the physical plausibility and origin of such a mass sheet\footnote{The \textit{internal} mass-sheet does not have to be a physical sheet of mass in the strictest terms, but could also reflect minor deviations from the actual input mass model.}. But, our analysis shows that upper limits on its spatial extent can be provided by means of \jwst-like stellar kinematics, unless $r_{\rm c} \gtrapprox 100$\,kpc. This will be helpful in further constraining physical models that are put forward for producing cored density distributions. Given this forecast, much of the precision lost by incorporating a maximally degenerate model can be regained, such that only a few systems with similar precision will be sufficient to address the $H_0$ tension between late and early time cosmological probes.
As new IFU facilities come online in the next years, we advocate for utilising this unique combination of lensing and dynamics, particularly with spatially resolved kinematics, as a powerful probe of cosmology.

%============================= Acknowledgements =============================
\section*{Acknowledgements}

The authors thank S.~Birrer, F.~Courbin, D.~Sluse and T.~Treu for helpful discussions and comments on the manuscript. AY and SHS thank the Max Planck Society for support through the Max Planck Research Group for SHS. G C.-F. C acknowledges support from the National Science Foundation through grants NSF-AST-1906976 and NSF-AST-1836016. G. C.-F. C. acknowledges support from the Moore Foundation through grant 8548. SHS and EK are supported in part by the Deutsche Forschungsgemeinschaft (DFG, German Research Foundation) under Germany's Excellence Strategy - EXC-2094 - 390783311. The Kavli IPMU is supported by World Premier International Research Center Initiative (WPI), MEXT, Japan.

%-------------------------------------------------------------
%                   For appendices and landscape, large table:
%                    in the preamble, use: \usepackage{lscape}
%-------------------------------------------------------------
%============================= References =============================
\bibliographystyle{aa}
\bibliography{aa}

\begin{thebibliography}{65}
\expandafter\ifx\csname natexlab\endcsname\relax\def\natexlab#1{#1}\fi

\bibitem[{{Abbott} {et~al.}(2018){Abbott}, {Abdalla}, {Annis}, {Bechtol},
  {Blazek}, {Benson}, {Bernstein}, {Bernstein}, {Bertin}, {Brooks}, {Burke},
  {Carnero Rosell}, {Carrasco Kind}, {Carretero}, {Castander}, {Chang},
  {Crawford}, {Cunha}, {D'Andrea}, {da Costa}, {Davis}, {DeRose}, {Desai},
  {Diehl}, {Dietrich}, {Doel}, {Drlica-Wagner}, {Evrard}, {Fernand ez},
  {Flaugher}, {Fosalba}, {Frieman}, {Garc{\'\i}a-Bellido}, {Gaztanaga},
  {Gerdes}, {Giannantonio}, {Gruen}, {Gruendl}, {Gschwend}, {Gutierrez},
  {Hartley}, {Henning}, {Honscheid}, {Hoyle}, {Huterer}, {Jain}, {James},
  {Jarvis}, {Jeltema}, {Johnson}, {Johnson}, {Krause}, {Kuehn}, {Kuhlmann},
  {Kuropatkin}, {Lahav}, {Liddle}, {Lima}, {Lin}, {MacCrann}, {Maia},
  {Manzotti}, {March}, {Marshall}, {Miquel}, {Mohr}, {Natoli}, {Nugent},
  {Ogando}, {Park}, {Plazas}, {Reichardt}, {Reil}, {Roodman}, {Ross}, {Rozo},
  {Rykoff}, {Sanchez}, {Scarpine}, {Schubnell}, {Scolnic}, {Sevilla-Noarbe},
  {Sheldon}, {Smith}, {Smith}, {Soares-Santos}, {Sobreira}, {Suchyta}, {Tarle},
  {Thomas}, {Troxel}, {Walker}, {Wechsler}, {Weller}, {Wester}, {Wu}, {Zuntz},
  {Dark Energy Survey Collaboration}, \& {South Pole Telescope
  Collaboration}}]{2018MNRAS.480.3879A}
{Abbott}, T.~M.~C., {Abdalla}, F.~B., {Annis}, J., {et~al.} 2018, \mnras, 480,
  3879

\bibitem[{{Agrawal} {et~al.}(2019){Agrawal}, {Cyr-Racine}, {Pinner}, \&
  {Randall}}]{2019arXiv190401016A}
{Agrawal}, P., {Cyr-Racine}, F.-Y., {Pinner}, D., \& {Randall}, L. 2019, arXiv
  e-prints, arXiv:1904.01016

\bibitem[{{Barkana}(1998)}]{1998ApJ...502..531B}
{Barkana}, R. 1998, \apj, 502, 531

\bibitem[{{Barnab{\`e}} {et~al.}(2012){Barnab{\`e}}, {Dutton}, {Marshall},
  {Auger}, {Brewer}, {Treu}, {Bolton}, {Koo}, \&
  {Koopmans}}]{2012MNRAS.423.1073B}
{Barnab{\`e}}, M., {Dutton}, A.~A., {Marshall}, P.~J., {et~al.} 2012, \mnras,
  423, 1073

\bibitem[{{Binney} \& {Tremaine}(1987)}]{1987gady.book.....B}
{Binney}, J. \& {Tremaine}, S. 1987, {Galactic Dynamics}

\bibitem[{{Birrer} {et~al.}(2020){Birrer}, {Shajib}, {Galan}, {Millon}, {Treu},
  {Agnello}, {Auger}, {Chen}, {Christensen}, {Collett}, {Courbin}, {Fassnacht},
  {Koopmans}, {Marshall}, {Park}, {Rusu}, {Sluse}, {Spiniello}, {Suyu},
  {Wagner-Carena}, {Wong}, {Barnab{\`e}}, {Bolton}, {Czoske}, {Ding},
  {Frieman}, \& {Van de Vyvere}}]{2020A&A...643A.165B}
{Birrer}, S., {Shajib}, A.~J., {Galan}, A., {et~al.} 2020, \aap, 643, A165

\bibitem[{{Birrer} \& {Treu}(2021)}]{2021A&A...649A..61B}
{Birrer}, S. \& {Treu}, T. 2021, \aap, 649, A61

\bibitem[{{Birrer} {et~al.}(2019){Birrer}, {Treu}, {Rusu}, {Bonvin},
  {Fassnacht}, {Chan}, {Agnello}, {Shajib}, {Chen}, {Auger}, {Courbin},
  {Hilbert}, {Sluse}, {Suyu}, {Wong}, {Marshall}, {Lemaux}, \&
  {Meylan}}]{2019MNRAS.484.4726B}
{Birrer}, S., {Treu}, T., {Rusu}, C.~E., {et~al.} 2019, \mnras, 484, 4726

\bibitem[{{Blum} {et~al.}(2020){Blum}, {Castorina}, \&
  {Simonovi{\'c}}}]{2020ApJ...892L..27B}
{Blum}, K., {Castorina}, E., \& {Simonovi{\'c}}, M. 2020, \apjl, 892, L27

\bibitem[{{Blum} \& {Teodori}(2021)}]{2021arXiv210510873B}
{Blum}, K. \& {Teodori}, L. 2021, arXiv e-prints, arXiv:2105.10873

\bibitem[{{Cappellari}(2002)}]{2002MNRAS.333..400C}
{Cappellari}, M. 2002, \mnras, 333, 400

\bibitem[{{Cappellari}(2008)}]{2008MNRAS.390...71C}
{Cappellari}, M. 2008, \mnras, 390, 71

\bibitem[{{Cappellari}(2020)}]{2020MNRAS.494.4819C}
{Cappellari}, M. 2020, \mnras, 494, 4819

\bibitem[{{Cappellari} \& {Copin}(2003)}]{2003MNRAS.342..345C}
{Cappellari}, M. \& {Copin}, Y. 2003, \mnras, 342, 345

\bibitem[{{Cappellari} {et~al.}(2007){Cappellari}, {Emsellem}, {Bacon},
  {Bureau}, {Davies}, {de Zeeuw}, {Falc{\'o}n-Barroso}, {Krajnovi{\'c}},
  {Kuntschner}, {McDermid}, {Peletier}, {Sarzi}, {van den Bosch}, \& {van de
  Ven}}]{2007MNRAS.379..418C}
{Cappellari}, M., {Emsellem}, E., {Bacon}, R., {et~al.} 2007, \mnras, 379, 418

\bibitem[{{Cappellari} {et~al.}(2013){Cappellari}, {Scott}, {Alatalo}, {Blitz},
  {Bois}, {Bournaud}, {Bureau}, {Crocker}, {Davies}, {Davis}, {de Zeeuw},
  {Duc}, {Emsellem}, {Khochfar}, {Krajnovi{\'c}}, {Kuntschner}, {McDermid},
  {Morganti}, {Naab}, {Oosterloo}, {Sarzi}, {Serra}, {Weijmans}, \&
  {Young}}]{2013MNRAS.432.1709C}
{Cappellari}, M., {Scott}, N., {Alatalo}, K., {et~al.} 2013, \mnras, 432, 1709

\bibitem[{{Chen} {et~al.}(2019){Chen}, {Fassnacht}, {Suyu}, {Rusu}, {Chan},
  {Wong}, {Auger}, {Hilbert}, {Bonvin}, {Birrer}, {Millon}, {Koopmans},
  {Lagattuta}, {McKean}, {Vegetti}, {Courbin}, {Ding}, {Halkola}, {Jee},
  {Shajib}, {Sluse}, {Sonnenfeld}, \& {Treu}}]{2019MNRAS.490.1743C}
{Chen}, G. C.~F., {Fassnacht}, C.~D., {Suyu}, S.~H., {et~al.} 2019, \mnras,
  490, 1743

\bibitem[{{Chen} {et~al.}(2020){Chen}, {Fassnacht}, {Suyu},
  {Y{\i}ld{\i}r{\i}m}, {Komatsu}, \& {Bernal}}]{2020arXiv201106002C}
{Chen}, G. C.~F., {Fassnacht}, C.~D., {Suyu}, S.~H., {et~al.} 2020, arXiv
  e-prints, arXiv:2011.06002

\bibitem[{{Emsellem} {et~al.}(1994){Emsellem}, {Monnet}, \&
  {Bacon}}]{1994A&A...285..723E}
{Emsellem}, E., {Monnet}, G., \& {Bacon}, R. 1994, \aap, 285, 723

\bibitem[{{Falco} {et~al.}(1985){Falco}, {Gorenstein}, \&
  {Shapiro}}]{1985ApJ...289L...1F}
{Falco}, E.~E., {Gorenstein}, M.~V., \& {Shapiro}, I.~I. 1985, \apjl, 289, L1

\bibitem[{{Foreman-Mackey} {et~al.}(2013){Foreman-Mackey}, {Hogg}, {Lang}, \&
  {Goodman}}]{2013PASP..125..306F}
{Foreman-Mackey}, D., {Hogg}, D.~W., {Lang}, D., \& {Goodman}, J. 2013, \pasp,
  125, 306

\bibitem[{{Freedman} {et~al.}(2019){Freedman}, {Madore}, {Hatt}, {Hoyt},
  {Jang}, {Beaton}, {Burns}, {Lee}, {Monson}, {Neeley}, {Phillips}, {Rich}, \&
  {Seibert}}]{2019ApJ...882...34F}
{Freedman}, W.~L., {Madore}, B.~F., {Hatt}, D., {et~al.} 2019, \apj, 882, 34

\bibitem[{{Gerhard}(1993)}]{1993MNRAS.265..213G}
{Gerhard}, O.~E. 1993, \mnras, 265, 213

\bibitem[{{Jeans}(1922)}]{1922MNRAS..82..122J}
{Jeans}, J.~H. 1922, \mnras, 82, 122

\bibitem[{{Jee} {et~al.}(2015){Jee}, {Komatsu}, \&
  {Suyu}}]{2015JCAP...11..033J}
{Jee}, I., {Komatsu}, E., \& {Suyu}, S.~H. 2015, \jcap, 11, 033

\bibitem[{{Jee} {et~al.}(2016){Jee}, {Komatsu}, {Suyu}, \&
  {Huterer}}]{2016JCAP...04..031J}
{Jee}, I., {Komatsu}, E., {Suyu}, S.~H., \& {Huterer}, D. 2016, \jcap, 4, 031

\bibitem[{{Jee} {et~al.}(2019){Jee}, {Suyu}, {Komatsu}, {Fassnacht}, {Hilbert},
  \& {Koopmans}}]{2019Sci...365.1134J}
{Jee}, I., {Suyu}, S.~H., {Komatsu}, E., {et~al.} 2019, Science, 365, 1134

\bibitem[{{Kassiola} \& {Kovner}(1993)}]{Kassiola+93}
{Kassiola}, A. \& {Kovner}, I. 1993, \apj, 417, 450

\bibitem[{{Knox} \& {Millea}(2020)}]{2020PhRvD.101d3533K}
{Knox}, L. \& {Millea}, M. 2020, \prd, 101, 043533

\bibitem[{{Kochanek}(2020)}]{2020MNRAS.493.1725K}
{Kochanek}, C.~S. 2020, \mnras, 493, 1725

\bibitem[{{Millon} {et~al.}(2020{\natexlab{a}}){Millon}, {Courbin}, {Bonvin},
  {Paic}, {Meylan}, {Tewes}, {Sluse}, {Magain}, {Chan}, {Galan}, {Joseph},
  {Lemon}, {Tihhonova}, {Anderson}, {Marmier}, {Chazelas}, {Lendl}, {Triaud},
  \& {Wyttenbach}}]{2020A&A...640A.105M}
{Millon}, M., {Courbin}, F., {Bonvin}, V., {et~al.} 2020{\natexlab{a}}, \aap,
  640, A105

\bibitem[{{Millon} {et~al.}(2020{\natexlab{b}}){Millon}, {Galan}, {Courbin},
  {Treu}, {Suyu}, {Ding}, {Birrer}, {Chen}, {Shajib}, {Sluse}, {Wong},
  {Agnello}, {Auger}, {Buckley-Geer}, {Chan}, {Collett}, {Fassnacht},
  {Hilbert}, {Koopmans}, {Motta}, {Mukherjee}, {Rusu}, {Sonnenfeld},
  {Spiniello}, \& {Van de Vyvere}}]{2020A&A...639A.101M}
{Millon}, M., {Galan}, A., {Courbin}, F., {et~al.} 2020{\natexlab{b}}, \aap,
  639, A101

\bibitem[{{Navarro} {et~al.}(1996){Navarro}, {Frenk}, \&
  {White}}]{1996ApJ...462..563N}
{Navarro}, J.~F., {Frenk}, C.~S., \& {White}, S.~D.~M. 1996, \apj, 462, 563

\bibitem[{{Navarro} {et~al.}(1997){Navarro}, {Frenk}, \&
  {White}}]{1997ApJ...490..493N}
{Navarro}, J.~F., {Frenk}, C.~S., \& {White}, S.~D.~M. 1997, \apj, 490, 493

\bibitem[{{Paraficz} \& {Hjorth}(2009)}]{2009A&A...507L..49P}
{Paraficz}, D. \& {Hjorth}, J. 2009, \aap, 507, L49

\bibitem[{{Pesce} {et~al.}(2020){Pesce}, {Braatz}, {Reid}, {Riess}, {Scolnic},
  {Condon}, {Gao}, {Henkel}, {Impellizzeri}, {Kuo}, \&
  {Lo}}]{2020ApJ...891L...1P}
{Pesce}, D.~W., {Braatz}, J.~A., {Reid}, M.~J., {et~al.} 2020, \apjl, 891, L1

\bibitem[{{Planck Collaboration} {et~al.}(2020){Planck Collaboration},
  {Aghanim}, {Akrami}, {Ashdown}, {Aumont}, {Baccigalupi}, {Ballardini},
  {Banday}, {Barreiro}, {Bartolo}, {Basak}, {Battye}, {Benabed}, {Bernard},
  {Bersanelli}, {Bielewicz}, {Bock}, {Bond}, {Borrill}, {Bouchet}, {Boulanger},
  {Bucher}, {Burigana}, {Butler}, {Calabrese}, {Cardoso}, {Carron},
  {Challinor}, {Chiang}, {Chluba}, {Colombo}, {Combet}, {Contreras}, {Crill},
  {Cuttaia}, {de Bernardis}, {de Zotti}, {Delabrouille}, {Delouis}, {Di
  Valentino}, {Diego}, {Dor{\'e}}, {Douspis}, {Ducout}, {Dupac}, {Dusini},
  {Efstathiou}, {Elsner}, {En{\ss}lin}, {Eriksen}, {Fantaye}, {Farhang},
  {Fergusson}, {Fernandez-Cobos}, {Finelli}, {Forastieri}, {Frailis},
  {Fraisse}, {Franceschi}, {Frolov}, {Galeotta}, {Galli}, {Ganga},
  {G{\'e}nova-Santos}, {Gerbino}, {Ghosh}, {Gonz{\'a}lez-Nuevo}, {G{\'o}rski},
  {Gratton}, {Gruppuso}, {Gudmundsson}, {Hamann}, {Handley}, {Hansen},
  {Herranz}, {Hildebrandt}, {Hivon}, {Huang}, {Jaffe}, {Jones}, {Karakci},
  {Keih{\"a}nen}, {Keskitalo}, {Kiiveri}, {Kim}, {Kisner}, {Knox},
  {Krachmalnicoff}, {Kunz}, {Kurki-Suonio}, {Lagache}, {Lamarre}, {Lasenby},
  {Lattanzi}, {Lawrence}, {Le Jeune}, {Lemos}, {Lesgourgues}, {Levrier},
  {Lewis}, {Liguori}, {Lilje}, {Lilley}, {Lindholm}, {L{\'o}pez-Caniego},
  {Lubin}, {Ma}, {Mac{\'\i}as-P{\'e}rez}, {Maggio}, {Maino}, {Mandolesi},
  {Mangilli}, {Marcos-Caballero}, {Maris}, {Martin}, {Martinelli},
  {Mart{\'\i}nez-Gonz{\'a}lez}, {Matarrese}, {Mauri}, {McEwen}, {Meinhold},
  {Melchiorri}, {Mennella}, {Migliaccio}, {Millea}, {Mitra},
  {Miville-Desch{\^e}nes}, {Molinari}, {Montier}, {Morgante}, {Moss}, {Natoli},
  {N{\o}rgaard-Nielsen}, {Pagano}, {Paoletti}, {Partridge}, {Patanchon},
  {Peiris}, {Perrotta}, {Pettorino}, {Piacentini}, {Polastri}, {Polenta},
  {Puget}, {Rachen}, {Reinecke}, {Remazeilles}, {Renzi}, {Rocha}, {Rosset},
  {Roudier}, {Rubi{\~n}o-Mart{\'\i}n}, {Ruiz-Granados}, {Salvati}, {Sandri},
  {Savelainen}, {Scott}, {Shellard}, {Sirignano}, {Sirri}, {Spencer},
  {Sunyaev}, {Suur-Uski}, {Tauber}, {Tavagnacco}, {Tenti}, {Toffolatti},
  {Tomasi}, {Trombetti}, {Valenziano}, {Valiviita}, {Van Tent}, {Vibert},
  {Vielva}, {Villa}, {Vittorio}, {Wandelt}, {Wehus}, {White}, {White},
  {Zacchei}, \& {Zonca}}]{2020A&A...641A...6P}
{Planck Collaboration}, {Aghanim}, N., {Akrami}, Y., {et~al.} 2020, \aap, 641,
  A6

\bibitem[{{Poulin} {et~al.}(2019){Poulin}, {Smith}, {Karwal}, \&
  {Kamionkowski}}]{2019PhRvL.122v1301P}
{Poulin}, V., {Smith}, T.~L., {Karwal}, T., \& {Kamionkowski}, M. 2019, \prl,
  122, 221301

\bibitem[{{Refsdal}(1964)}]{1964MNRAS.128..307R}
{Refsdal}, S. 1964, \mnras, 128, 307

\bibitem[{{Riess} {et~al.}(2019){Riess}, {Casertano}, {Yuan}, {Macri}, \&
  {Scolnic}}]{2019ApJ...876...85R}
{Riess}, A.~G., {Casertano}, S., {Yuan}, W., {Macri}, L.~M., \& {Scolnic}, D.
  2019, \apj, 876, 85

\bibitem[{{Rusu} {et~al.}(2020){Rusu}, {Wong}, {Bonvin}, {Sluse}, {Suyu},
  {Fassnacht}, {Chan}, {Hilbert}, {Auger}, {Sonnenfeld}, {Birrer}, {Courbin},
  {Treu}, {Chen}, {Halkola}, {Koopmans}, {Marshall}, \&
  {Shajib}}]{2020MNRAS.498.1440R}
{Rusu}, C.~E., {Wong}, K.~C., {Bonvin}, V., {et~al.} 2020, \mnras, 498, 1440

\bibitem[{{Schneider} \& {Sluse}(2013)}]{2013A&A...559A..37S}
{Schneider}, P. \& {Sluse}, D. 2013, \aap, 559, A37

\bibitem[{{Schneider} \& {Sluse}(2014)}]{2014A&A...564A.103S}
{Schneider}, P. \& {Sluse}, D. 2014, \aap, 564, A103

\bibitem[{Schwarz(1978)}]{schwarz1978}
Schwarz, G. 1978, Ann. Statist., 6, 461

\bibitem[{{Shajib} {et~al.}(2020){Shajib}, {Birrer}, {Treu}, {Agnello},
  {Buckley-Geer}, {Chan}, {Christensen}, {Lemon}, {Lin}, {Millon}, {Poh},
  {Rusu}, {Sluse}, {Spiniello}, {Chen}, {Collett}, {Courbin}, {Fassnacht},
  {Frieman}, {Galan}, {Gilman}, {More}, {Anguita}, {Auger}, {Bonvin},
  {McMahon}, {Meylan}, {Wong}, {Abbott}, {Annis}, {Avila}, {Bechtol}, {Brooks},
  {Brout}, {Burke}, {Carnero Rosell}, {Carrasco Kind}, {Carretero},
  {Castander}, {Costanzi}, {da Costa}, {De Vicente}, {Desai}, {Dietrich},
  {Doel}, {Drlica-Wagner}, {Evrard}, {Finley}, {Flaugher}, {Fosalba},
  {Garc{\'\i}a-Bellido}, {Gerdes}, {Gruen}, {Gruendl}, {Gschwend}, {Gutierrez},
  {Hollowood}, {Honscheid}, {Huterer}, {James}, {Jeltema}, {Krause},
  {Kuropatkin}, {Li}, {Lima}, {MacCrann}, {Maia}, {Marshall}, {Melchior},
  {Miquel}, {Ogando}, {Palmese}, {Paz-Chinch{\'o}n}, {Plazas}, {Romer},
  {Roodman}, {Sako}, {Sanchez}, {Santiago}, {Scarpine}, {Schubnell}, {Scolnic},
  {Serrano}, {Sevilla-Noarbe}, {Smith}, {Soares-Santos}, {Suchyta}, {Tarle},
  {Thomas}, {Walker}, \& {Zhang}}]{2020MNRAS.494.6072S}
{Shajib}, A.~J., {Birrer}, S., {Treu}, T., {et~al.} 2020, \mnras, 494, 6072

\bibitem[{{Shanks} {et~al.}(2019){Shanks}, {Hogarth}, \&
  {Metcalfe}}]{2019MNRAS.484L..64S}
{Shanks}, T., {Hogarth}, L.~M., \& {Metcalfe}, N. 2019, \mnras, 484, L64

\bibitem[{{Sluse} {et~al.}(2003){Sluse}, {Surdej}, {Claeskens},
  {Hutsem{\'e}kers}, {Jean}, {Courbin}, {Nakos}, {Billeres}, \&
  {Khmil}}]{2003A&A...406L..43S}
{Sluse}, D., {Surdej}, J., {Claeskens}, J.-F., {et~al.} 2003, \aap, 406, L43

\bibitem[{{Sonnenfeld}(2018)}]{2018MNRAS.474.4648S}
{Sonnenfeld}, A. 2018, \mnras, 474, 4648

\bibitem[{{Suyu} {et~al.}(2013){Suyu}, {Auger}, {Hilbert}, {Marshall}, {Tewes},
  {Treu}, {Fassnacht}, {Koopmans}, {Sluse}, {Blandford}, {Courbin}, \&
  {Meylan}}]{2013ApJ...766...70S}
{Suyu}, S.~H., {Auger}, M.~W., {Hilbert}, S., {et~al.} 2013, \apj, 766, 70

\bibitem[{{Suyu} {et~al.}(2017){Suyu}, {Bonvin}, {Courbin}, {Fassnacht},
  {Rusu}, {Sluse}, {Treu}, {Wong}, {Auger}, {Ding}, {Hilbert}, {Marshall},
  {Rumbaugh}, {Sonnenfeld}, {Tewes}, {Tihhonova}, {Agnello}, {Blandford},
  {Chen}, {Collett}, {Koopmans}, {Liao}, {Meylan}, \&
  {Spiniello}}]{2017MNRAS.468.2590S}
{Suyu}, S.~H., {Bonvin}, V., {Courbin}, F., {et~al.} 2017, \mnras, 468, 2590

\bibitem[{{Suyu} {et~al.}(2018){Suyu}, {Chang}, {Courbin}, \&
  {Okumura}}]{2018SSRv..214...91S}
{Suyu}, S.~H., {Chang}, T.-C., {Courbin}, F., \& {Okumura}, T. 2018, \ssr, 214,
  91

\bibitem[{{Suyu} {et~al.}(2010){Suyu}, {Marshall}, {Auger}, {Hilbert},
  {Blandford}, {Koopmans}, {Fassnacht}, \& {Treu}}]{2010ApJ...711..201S}
{Suyu}, S.~H., {Marshall}, P.~J., {Auger}, M.~W., {et~al.} 2010, \apj, 711, 201

\bibitem[{{Suyu} {et~al.}(2006){Suyu}, {Marshall}, {Hobson}, \&
  {Blandford}}]{2006MNRAS.371..983S}
{Suyu}, S.~H., {Marshall}, P.~J., {Hobson}, M.~P., \& {Blandford}, R.~D. 2006,
  \mnras, 371, 983

\bibitem[{{Suyu} {et~al.}(2014){Suyu}, {Treu}, {Hilbert}, {Sonnenfeld},
  {Auger}, {Blandford}, {Collett}, {Courbin}, {Fassnacht}, {Koopmans},
  {Marshall}, {Meylan}, {Spiniello}, \& {Tewes}}]{2014ApJ...788L..35S}
{Suyu}, S.~H., {Treu}, T., {Hilbert}, S., {et~al.} 2014, \apjl, 788, L35

\bibitem[{{Tewes} {et~al.}(2013{\natexlab{a}}){Tewes}, {Courbin}, \&
  {Meylan}}]{2013A&A...553A.120T}
{Tewes}, M., {Courbin}, F., \& {Meylan}, G. 2013{\natexlab{a}}, \aap, 553, A120

\bibitem[{{Tewes} {et~al.}(2013{\natexlab{b}}){Tewes}, {Courbin}, {Meylan},
  {Kochanek}, {Eulaers}, {Cantale}, {Mosquera}, {Magain}, {Van Winckel},
  {Sluse}, {Cataldi}, {V{\"o}r{\"o}s}, \& {Dye}}]{2013A&A...556A..22T}
{Tewes}, M., {Courbin}, F., {Meylan}, G., {et~al.} 2013{\natexlab{b}}, \aap,
  556, A22

\bibitem[{{Treu} \& {Marshall}(2016)}]{2016A&ARv..24...11T}
{Treu}, T. \& {Marshall}, P.~J. 2016, \aapr, 24, 11

\bibitem[{{Unruh} {et~al.}(2017){Unruh}, {Schneider}, \&
  {Sluse}}]{2017A&A...601A..77U}
{Unruh}, S., {Schneider}, P., \& {Sluse}, D. 2017, \aap, 601, A77

\bibitem[{{van de Ven} {et~al.}(2010){van de Ven}, {Falc{\'o}n-Barroso},
  {McDermid}, {Cappellari}, {Miller}, \& {de Zeeuw}}]{2010ApJ...719.1481V}
{van de Ven}, G., {Falc{\'o}n-Barroso}, J., {McDermid}, R.~M., {et~al.} 2010,
  \apj, 719, 1481

\bibitem[{{van der Marel} \& {Franx}(1993)}]{1993ApJ...407..525V}
{van der Marel}, R.~P. \& {Franx}, M. 1993, \apj, 407, 525

\bibitem[{{Verde} {et~al.}(2019){Verde}, {Treu}, \&
  {Riess}}]{2019NatAs...3..891V}
{Verde}, L., {Treu}, T., \& {Riess}, A.~G. 2019, Nature Astronomy, 3, 891

\bibitem[{{Wertz} {et~al.}(2018){Wertz}, {Orthen}, \&
  {Schneider}}]{2018A&A...617A.140W}
{Wertz}, O., {Orthen}, B., \& {Schneider}, P. 2018, \aap, 617, A140

\bibitem[{{Wong} {et~al.}(2017){Wong}, {Suyu}, {Auger}, {Bonvin}, {Courbin},
  {Fassnacht}, {Halkola}, {Rusu}, {Sluse}, {Sonnenfeld}, {Treu}, {Collett},
  {Hilbert}, {Koopmans}, {Marshall}, \& {Rumbaugh}}]{2017MNRAS.465.4895W}
{Wong}, K.~C., {Suyu}, S.~H., {Auger}, M.~W., {et~al.} 2017, \mnras, 465, 4895

\bibitem[{{Wong} {et~al.}(2020){Wong}, {Suyu}, {Chen}, {Rusu}, {Millon},
  {Sluse}, {Bonvin}, {Fassnacht}, {Taubenberger}, {Auger}, {Birrer}, {Chan},
  {Courbin}, {Hilbert}, {Tihhonova}, {Treu}, {Agnello}, {Ding}, {Jee},
  {Komatsu}, {Shajib}, {Sonnenfeld}, {Blandford}, {Koopmans}, {Marshall}, \&
  {Meylan}}]{2020MNRAS.498.1420W}
{Wong}, K.~C., {Suyu}, S.~H., {Chen}, G.~C.~F., {et~al.} 2020, \mnras, 498,
  1420

\bibitem[{{Y{\i}ld{\i}r{\i}m} {et~al.}(2020){Y{\i}ld{\i}r{\i}m}, {Suyu}, \&
  {Halkola}}]{2020MNRAS.493.4783Y}
{Y{\i}ld{\i}r{\i}m}, A., {Suyu}, S.~H., \& {Halkola}, A. 2020, \mnras, 493,
  4783

\end{thebibliography}
%=====================================================================

%============================= Appendix =============================
\appendix
\section{}
\label{sec:appendixa}
\begin{figure*}
\begin{center}
\includegraphics[width=0.9\linewidth]{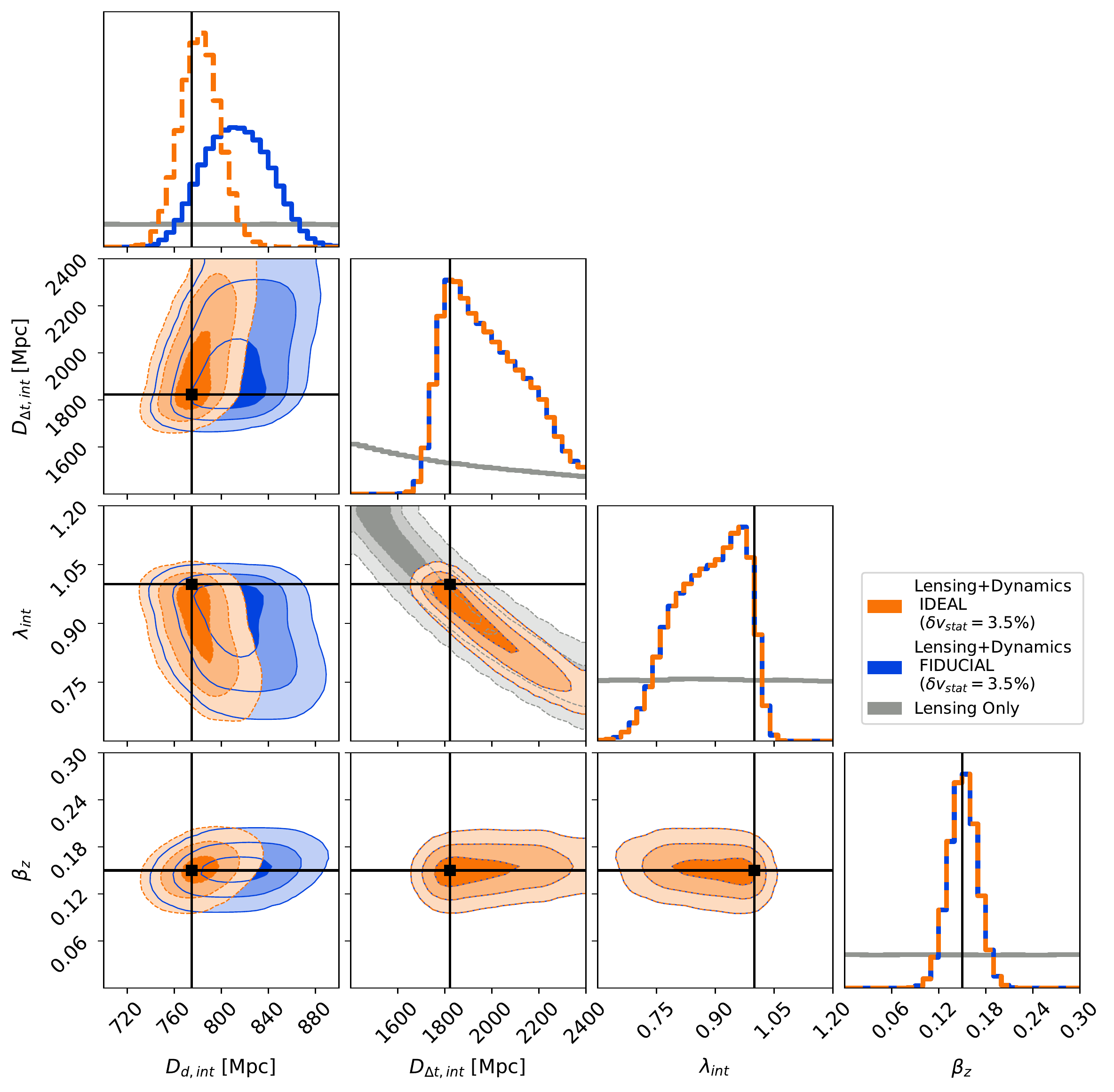}
\caption{Measurements from our joint strong lensing and stellar dynamical models, without any kinematic data (grey) and with mock IFU stellar kinematics (orange and blue) and an increased statistical error of $\delta v_{\rm stat} = 3.5\%$ for the kinematics. In concordance with Fig. \ref{fig:fig3}, the plot shows the constraints for the most relevant parameters in the fit, consisting of the lens distance \ddint, time-delay distance \dtint, internal mass-sheet parameter \lint\ and orbital anisotropy $\beta_z$. The contours highlight the 1-, 2- and 3$\sigma$ confidence regions. The black points (lines) depict the mock input values. The increased error of 3.5\% for each velocity bin translates into larger uncertainties on \dtint\ and \lint. The uncertainties on \ddint\ are dictated by the systematic offset ($\delta v_{\rm sys}$) and are unchanged with respect to the models in Sec. \ref{sec:results:distances}, given the same bias of $-2$\% in \vlosldata.}
\label{fig:fig_appendix_1}
\end{center}
\end{figure*}

\begin{figure*}
\begin{center}
\includegraphics[width=0.49\linewidth]{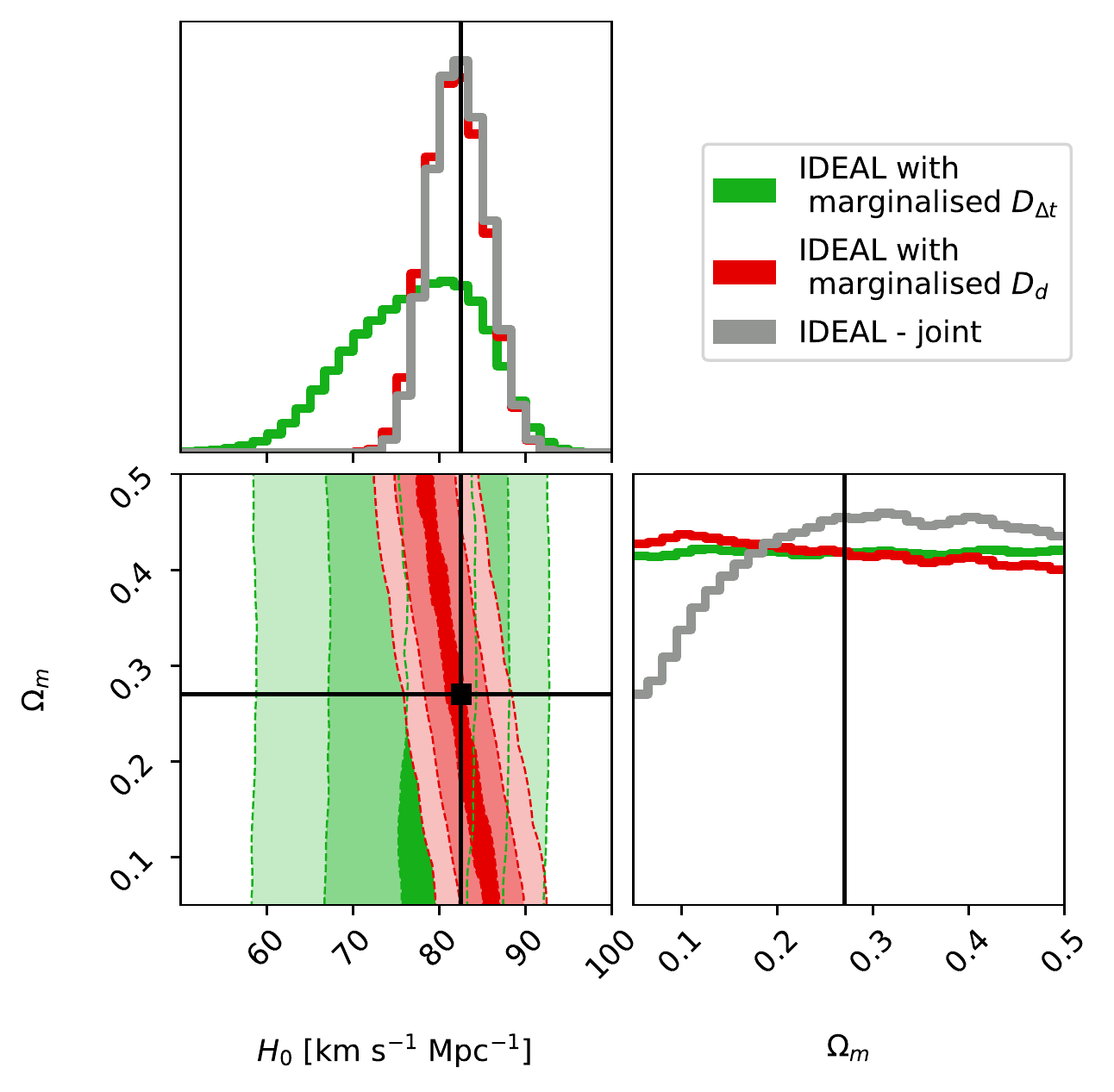}
\includegraphics[width=0.49\linewidth]{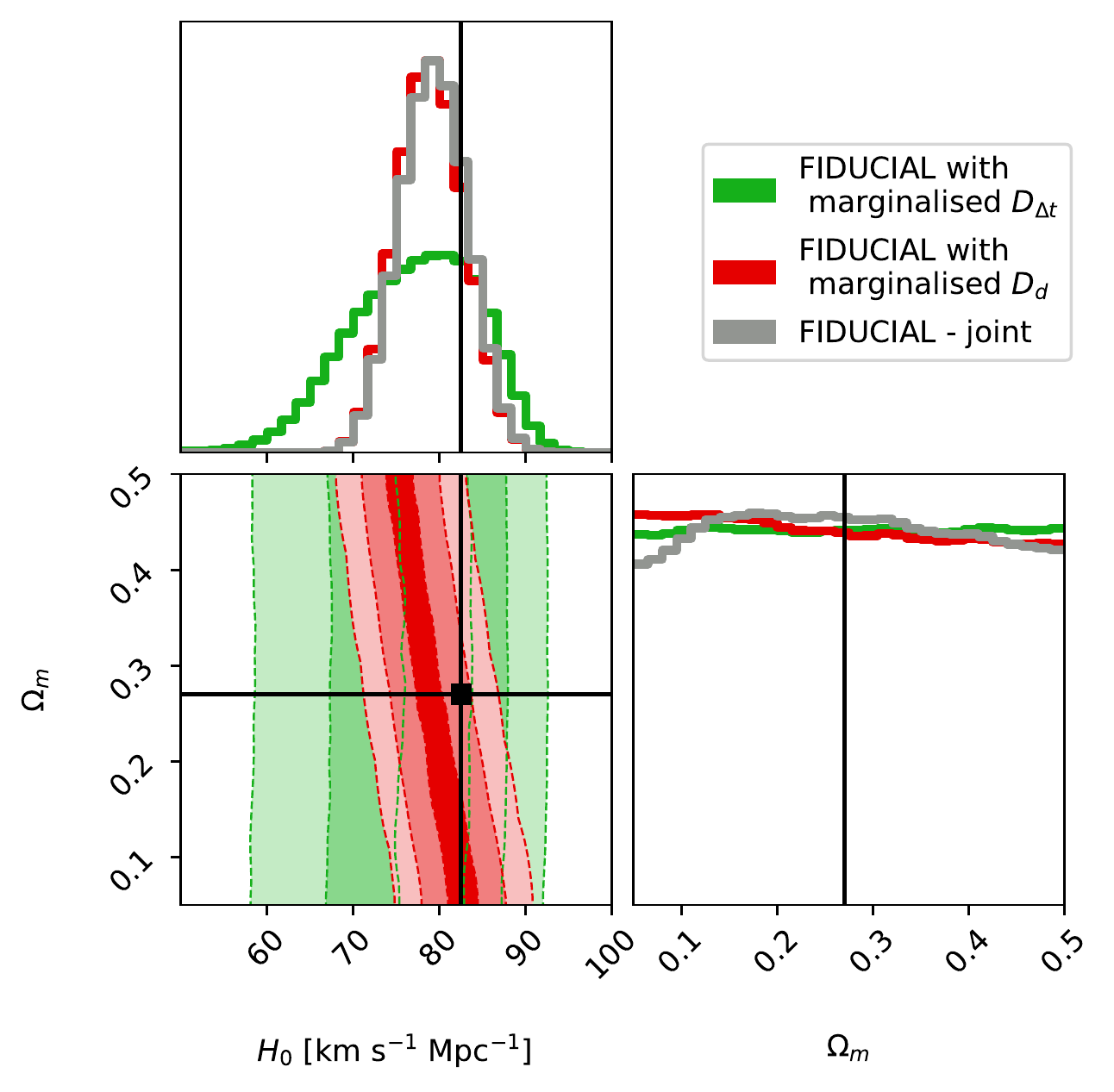}
\caption{\h\ and $\Omega_{\rm{m}}$ constraints from our models in flat \lcdm\ cosmology, for the IDEAL (left) and FIDUCIAL (data sets), with increased statistical errors of $\delta v_{\rm stat} = 3.5\%$. We adopted uniform priors on \h\ of [50, 120]\,\kmsM and on $\Omega_{\rm m}$ of [0.05, 0.5]. The green shaded regions show the 1, 2 and 3$\sigma$ confidence regions based on the constraint on $\dt$ only, i.e., $P(\dt)$ after marginalising over other mass model parameters including $\dd$. The red shaded regions show the corresponding constraints %marginalised over \dd. 
based on $P(\dd)$, after marginalising over $\dt$. The grey histogram shows the constraints based on $P(\dd,\dt)$, where we have omitted showing the 2D contours for ease of viewing.
In principle, the different tilts in the \h-$\Omega_{\rm m}$ space, when marginalised over \dd\ and \dt\ respectively, can  break some of the degeneracies. However, this requires highly accurate and precise measurements of the LOSVD. Errors of $\delta v_{\rm stat} = 3.5\%$ are already insufficient to provide powerful constraints on e.g. $\Omega_{\rm m}$ on a single lens basis when accounting for the MST.}
%the dark matter density. 
%Note that given the smaller overlap of the \dd\ and \dt\ marginalised models in the FIDUCIAL data case yields to slightly tighter constraints on the joint \h\ inference.}
\label{fig:fig_appendix_2}
\end{center}
\end{figure*}

We present the distance constraints and cosmological forecast from our joint strong lensing and stellar dynamical models, based on the IDEAL and FIDUCIAL data. Unlike the results shown in Fig. \ref{fig:fig3} and Fig. \ref{fig:fig6}, the statistical errror ($\delta v_{\rm stat}$) in the mock kinematics (Eq. \ref{eqn:eqn1}) has been increased from 2.5\% to 3.5\%, while $\delta v_{\rm sys}$ remains $-2$\% for the FIDUCIAL data.  This roughly corresponds to a target S/N of 30 per bin. As before, we assume a uniform distribution of $f_{\rm v} \in [-0.02, 0.02]$. Fig. \ref{fig:fig_appendix_1} effectively illustrates how the increased statistical uncertainty in the kinematics translates into larger modelling uncertainties on \dtint\ and \lint, whereas the uncertainty on \ddint\ depends mainly on the fractional offset $f_{\rm v}$ distribution that accounts for the systematic uncertainties in the kinematic measurements.
%systematic offset ($\delta v_{\rm sys}$) in the measured velocity moments. 
Given the degraded mock kinematics, the uncertainties on \dtint\ (further amplified by the degeneracy between \dtint\ and \lint) become significant, such that most of the constraining power in the cosmological inference hinges on the accuracy and precision of \ddint\ (Fig.\ref{fig:fig_appendix_2}). This is most obvious when drawing a direct comparison between the IDEAL and FIDUCIAL cosmological forecast here. We obtain a final \h\ measurement of $82.2^{+3.0}_{-2.8}$ \kmsM\ in the IDEAL case, whereas the FIDUCIAL data yields $H_{0} = 79.3^{+3.6}_{-3.6}$ \kmsM. The shift in the mean value for the FIDUCIAL data is mainly caused by the significant drop in the constraining power of \dtint. With the systematic offset in the velocity moments for the FIDUCIAL case, the mean \h\ is also biased. This is in contrast to our findings without a mass sheet in Y20 (see Table 3) where, despite increased statistical errors in the kinematic data, both \dtint\ and \ddint\ contribute almost equally to an accurate and precise \h\ measurement. We summarise the results in Table \ref{table:table_appendix_1}.

\begin{table*}
	\caption{Parameter constraints and cosmological forecast from our lensing-only and joint strong lensing \&\ stellar dynamical models. The IFU stellar kinematics have been mocked with (FIDUCIAL) and without (IDEAL) systematic errors included, but an increased statistical error of $\delta v_{\rm stat} = 3.5\%$. The last column represents the inferred \h\ values for the various models, under the assumption of flat \lcdm. The mock input value is 82.5 \kmsM. We quote the 50th and 16th/84th percentiles of the distribution.}
	\begin{center}
	\centerline{
	\begin{tabular}{ c | c  c  c c }
		\hline
		Model & \ddint\ [Mpc] & \dtint\ [Mpc] & \lint & \h$\,$[\kmsM] (flat \lcdm) \\
		\hline
		Lensing & - & 1749$^{+897}_{-891}$ & 1.00$^{+0.50}_{-0.50}$ & 82.7$^{+24}_{-22}$ \\
		\\
		Lensing \& Dynamics & 782$^{+15}_{-15}$ & 1965$^{+217}_{-153}$ & 0.89$^{+0.09}_{-0.10}$ & 82.2$^{+3.0}_{-2.8}$ \\
		IDEAL with $\delta v_{\rm stat} = 3.5\%$\\
		\\
		Lensing \& Dynamics & 813$^{+28}_{-28}$ & 1965$^{+217}_{-153}$ & 0.89$^{+0.08}_{-0.10}$ & 79.3$^{+3.6}_{-3.6}$ \\
		FIDUCIAL with $\delta v_{\rm stat} = 3.5\%$\\
		\\
        % \hline
	\end{tabular}}
\end{center}
%	}
%	\vspace{2ex}
	\label{table:table_appendix_1}
%	\end{center}
\end{table*}

\section{}
\label{sec:appendixb}
\begin{figure*}
\begin{center}
\includegraphics[width=0.9\linewidth]{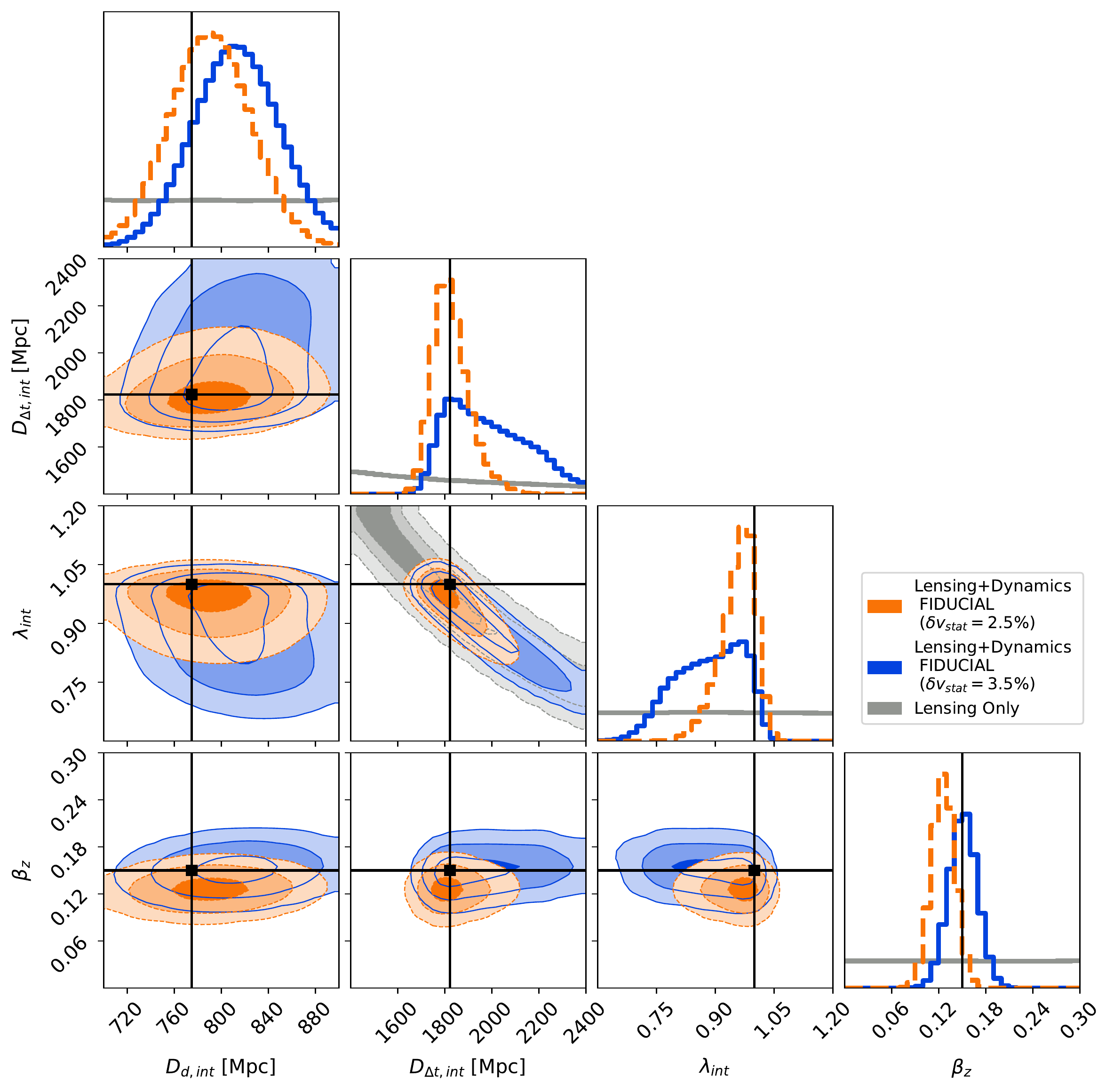}
\caption{Measurements from our joint strong lensing and stellar dynamical models, without any kinematic data (grey) and with mock IFU stellar kinematics (orange and blue). In both cases, we adopt $\delta v_{\rm sys} = -2\%$ and a statistical error of $\delta v_{\rm stat} = 2.5\%$ (orange) and $\delta v_{\rm stat} = 3.5\%$ (blue) respectively. In concordance with Fig. \ref{fig:fig3}, the plot shows the constraints for the most relevant parameters in the fit, consisting of the lens distance \ddint, time-delay distance \dtint, internal mass-sheet parameter \lint\ and orbital anisotropy $\beta_z$. The contours highlight the 1-, 2- and 3$\sigma$ confidence regions. The black points (lines) depict the mock input values. The increased error of 3.5\% for each velocity bin translates into larger uncertainties on \dtint\ and \lint. The uncertainties on \ddint\ are dictated by the systematic offset ($\delta v_{\rm sys}$) and are almost identical for both mock kinematic data sets, given the same bias of $-2$\% in \vlosldata.}
\label{fig:fig_appendix_4}
\end{center}
\end{figure*}

\begin{figure*}
\begin{center}
\includegraphics[width=0.49\linewidth]{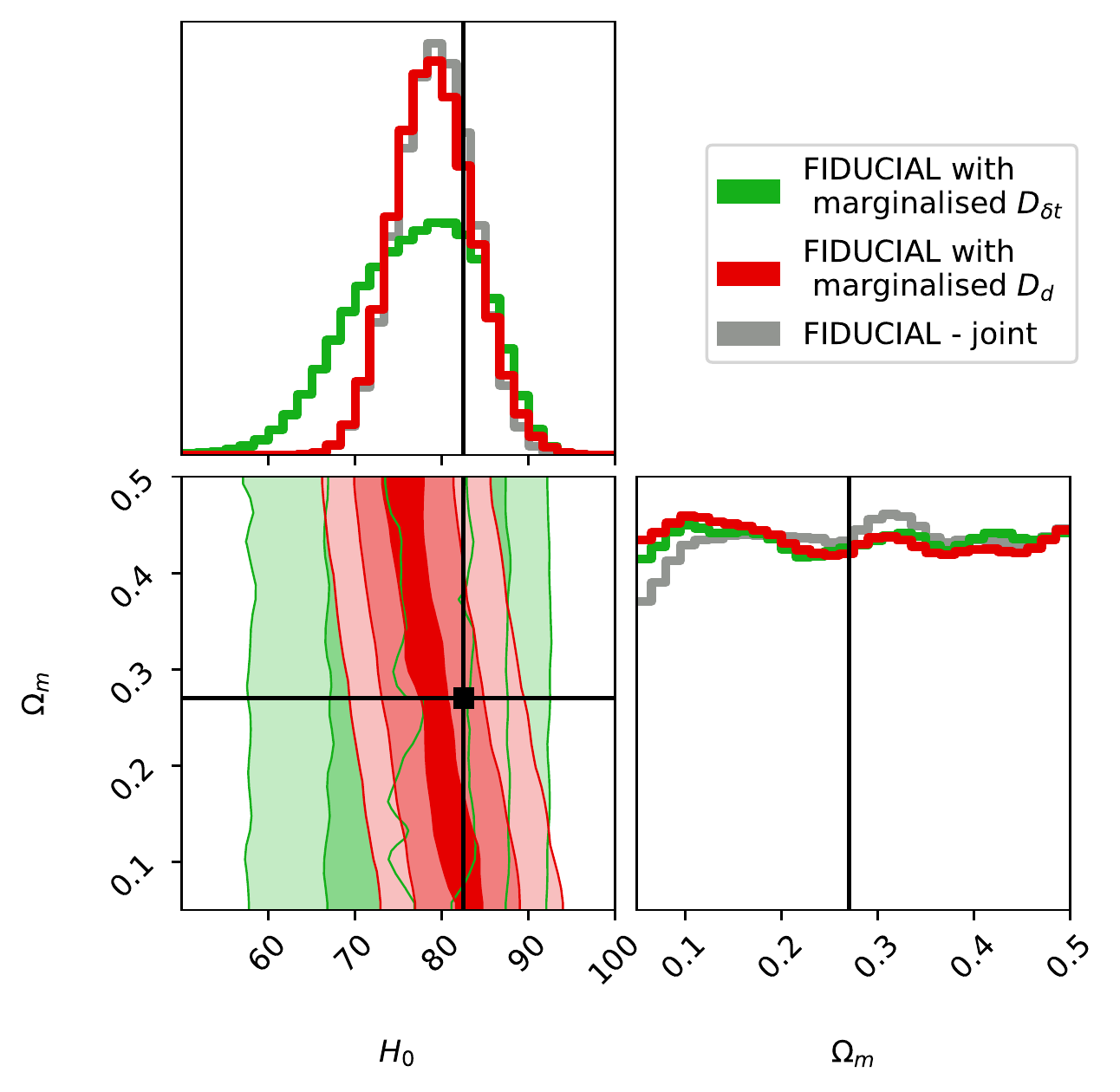}
\includegraphics[width=0.49\linewidth]{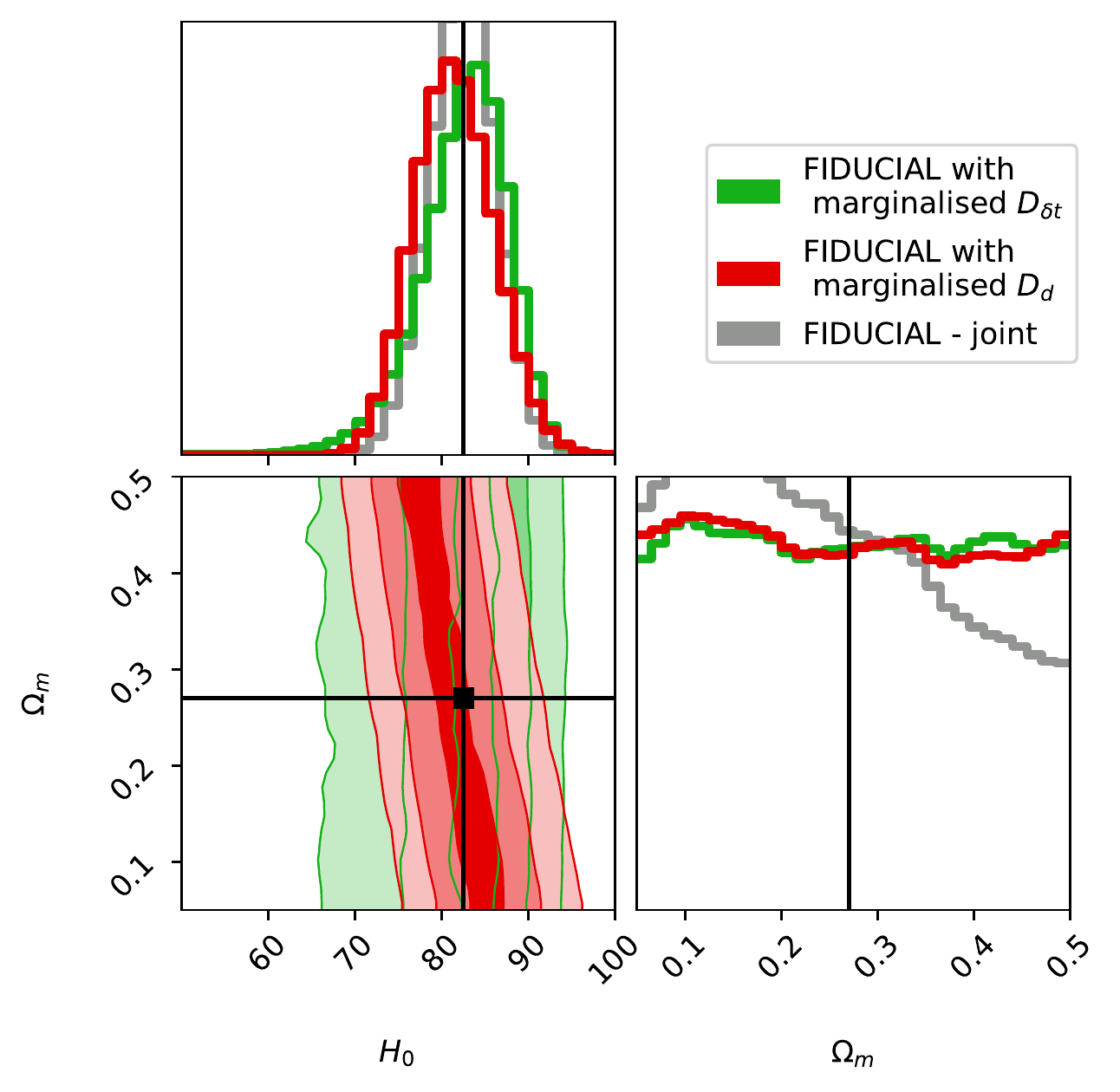}
\caption{\h\ and $\Omega_{\rm{m}}$ constraints from our models in flat \lcdm\ cosmology, for the FIDUCIAL data sets with $\delta v_{\rm stat} = 2.5\%$  (left) and $\delta v_{\rm stat} = 3.5\%$ (right). The systematic uncertainty is $\delta v_{\rm sys} = -2.0\%$, but assuming a Gaussian distribution for $f_{\rm v}$ within the range of [-0.06, 0.06] for both. We adopted uniform priors on \h\ of [50, 120]\,\kmsM and on $\Omega_{\rm m}$ of [0.05, 0.5]. The green shaded regions show the 1, 2 and 3$\sigma$ confidence regions based on the constraint on $\dt$ only, i.e., $P(\dt)$ after marginalising over other mass model parameters including $\dd$. The red shaded regions show the corresponding constraints %marginalised over \dd. 
based on $P(\dd)$, after marginalising over $\dt$. The grey histogram shows the constraints based on $P(\dd,\dt)$, where we have omitted showing the 2D contours for ease of viewing.
In principle, the different tilts in the \h-$\Omega_{\rm m}$ space, when marginalised over \dd\ and \dt\ respectively, can  break some of the degeneracies. However, this requires highly accurate and precise measurements of the LOSVD. Errors of $\delta v_{\rm stat} = 3.5\%$ are already insufficient to provide powerful constraints on e.g. $\Omega_{\rm m}$ on a single lens basis when accounting for the MST.}
%the dark matter density. 
%Note that given the smaller overlap of the \dd\ and \dt\ marginalised models in the FIDUCIAL data case yields to slightly tighter constraints on the joint \h\ inference.}
\label{fig:fig_appendix_5}
\end{center}
\end{figure*}

To facilitate a direct comparison with the cosmological forecast in \cite{2021A&A...649A..61B}, we perform the joint modelling of the strong lensing and stellar kinematic data with $\delta v_{\rm stat} = 2.5\%$ and $\delta v_{\rm sys} = 2\%$ (orange) and $\delta v_{\rm stat} = 3.5\%$ and $\delta v_{\rm sys} = 2\%$ (blue), and show the corresponding parameter constraints in Fig. \ref{fig:fig_appendix_4}. 
In contrast to Section \ref{sec:results:distances} where $f_{\rm v}$ was uniformly distributed between $[-0.02,0.02]$,
$f_{\rm v}$ is now assumed to be a Gaussian distribution with a standard deviation of 0.02, for which we create 25 realisations of $f_{\rm v}$, covering a range of [-0.06, 0.06] in steps of 0.005, and apply a weighting of each chain given by $w = \rm exp \big(-0.5\,(f_{\rm v}/\sigma_{\rm v})^{2} \big)$, with $\sigma_{\rm v} = 0.02$. The corresponding cosmological forecast for flat \lcdm\ is presented in Fig. \ref{fig:fig_appendix_5}, with the results summarised in Table \ref{table:table_appendix_2}.

\begin{table*}
	\caption{Parameter constraints and cosmological forecast from our lensing-only and joint strong lensing \&\ stellar dynamical models. The IFU stellar kinematics have been mocked with systematic errors included (FIDUCIAL), and two realisations for the statistical error of $\delta v_{\rm stat} = 2.5\%$ and $\delta v_{\rm stat} = 3.5\%$, respectively. The last column represents the inferred \h\ values for the various models, under the assumption of flat \lcdm. The mock input value is 82.5 \kmsM. We quote the 50th and 16th/84th percentiles of the distribution.}
	\begin{center}
	\centerline{
	\begin{tabular}{ c | c  c  c c }
		\hline
		Model & \ddint\ [Mpc] & \dtint\ [Mpc] & \lint & \h$\,$[\kmsM] (flat \lcdm) \\
		\hline
		Lensing & - & 1749$^{+897}_{-891}$ & 1.00$^{+0.50}_{-0.50}$ & 82.7$^{+24}_{-22}$ \\
		\\
		Lensing \& Dynamics & 791$^{+34}_{-34}$ & 1816$^{+81}_{-53}$ & 0.97$^{+0.03}_{-0.05}$ & 82.5$^{+3.2}_{-3.5}$ \\
		FIDUCIAL with $\delta v_{\rm stat} = 2.5\%$\\
		\\
		Lensing \& Dynamics & 813$^{+36}_{-36}$ & 1965$^{+217}_{-153}$ & 0.89$^{+0.08}_{-0.10}$ & 79.3$^{+4.0}_{-4.1}$ \\
		FIDUCIAL with $\delta v_{\rm stat} = 3.5\%$\\
		\\
        % \hline
	\end{tabular}}
\end{center}
%	}
%	\vspace{2ex}
	\label{table:table_appendix_2}
%	\end{center}
\end{table*}
%=====================================================================

\section{}
\label{sec:appendixc}

% \sherryII{\textbf{New Appendix}}

In this Appendix, we show in more detail that a MST with a large core radius for the 
cored surface mass-density profile (that approximates a mass sheet) effectively yields a vanishing density profile and does not influence the stellar kinematics of the lens galaxy.

We start from equation (\ref{eqn:eqn14}) for the projected SMD,
\begin{equation}
\label{eqn:AppC:SMD}
\begin{aligned}
\Sigma_{\rm smd}(\theta) & = \bigg[ \frac{c^{2}}{4\pi G}\;\frac{{\dtint}}{(1+\zd)\;\ddint^2}\bigg]\\
& \times \bigg[(1-\lint)+\lint\;\kappa_{\rm composite}(\theta) \bigg].\\
\end{aligned}
\end{equation}
The term $(1-\lint)$ represents the mass sheet, and following B20, we approximate it as a cored dimensionless surface-mass-density distribution\footnote{Our cored density profile has a different form compared to the one used in B20.},
\begin{equation}
\kappa_{\rm core}(\theta) = \frac{a_{0} \theta_{\rm c}}{\sqrt{\theta^2 + \theta_{\rm c}^2}} \simeq (1-\lint),
\label{eqn:AppC:kcore}
\end{equation}
where $a_0$ is a normalisation parameter that is directly proportional to the surface mass density of the mass sheet. For $\theta_{\rm c} \gg \theta$, the cored mass-density profile approximates well a mass sheet of constant convergence.  Assuming spherical symmetry, the mass density corresponding to the (projected) dimensionless surface-mass density above is
\begin{equation}
\rho_{\rm core}(r) =  \frac{a_0 r_{\rm c} \Sigma_{\mathrm{crit}} }{\pi (r^2 + r_{\rm c}^2)},  
\end{equation}
where $\Sigma_{\mathrm{crit}}$ is given by Equation (\ref{eqn:eqn13}).  
In other words, $\rho_{\rm core}$ is the deprojection of $\kappa_{\rm core}$.

Substituting equation (\ref{eqn:AppC:kcore}) into equation (\ref{eqn:AppC:SMD}), we obtain
\begin{equation}
\begin{aligned}
\Sigma_{\rm smd}(\theta) & = \bigg[ \frac{c^{2}}{4\pi G}\;\frac{{\dtint}}{(1+\zd)\;\ddint^2}\bigg] \bigg[\kappa_{\rm core} + \lint\;\kappa_{\rm composite}(\theta) \bigg].\\
\end{aligned}
\label{eq:AppC:SMD_core}
\end{equation}

In order to predict the $\overline{v_{\mathrm{LOS}}^{2}}$ through the Jeans equation, we would first deproject the SMD to obtain the mass density
\begin{equation}
\begin{aligned}
\rho(r) & = \bigg[ \frac{c^{2}}{4\pi G}\;\frac{{\dtint}}{(1+\zd)\;\ddint^2}\bigg]\\
& \times \bigg[\rho_{\rm core}(r) + \lint\;{\rm Deproj}\left[\kappa_{\rm composite}(\theta)\right] \bigg],\\
\end{aligned}
\end{equation}
where the deprojection for ${\rm Deproj} \left[\kappa_{\rm composite}(\theta)\right]$ can be done through the MGEs described in Section \ref{subsec:stellardynamics}.
For large $r_{\rm c}$ ($\gtrsim 400\,\rm{kpc}$ for our case), $\rho_{\rm core}(r) \rightarrow$  0 and we obtain 
\begin{equation}
\begin{aligned}
\rho(r) & = \frac{c^{2}}{4\pi G}\;\frac{{\dtint}\; \lint}{(1+\zd)\;\ddint^2} \;{\rm Deproj}\left[\kappa_{\rm composite}(\theta)\right].\\
\end{aligned}
\end{equation}
Substituting equation (\ref{eqn:eqn6}) into the above equation for $\dtint$, the $\lint$ in the numerator and denominator cancels and we find that
\begin{equation}
\begin{aligned}
\rho(r) & = \frac{c^{2}}{4\pi G}\;\frac{{D_{\Delta t, {\rm composite}}}}{(1+\zd)\;\ddint^2} \;{\rm Deproj}\left[\kappa_{\rm composite}(\theta)\right].\\
\end{aligned}
\end{equation}
Therefore, when $r_{\rm c}$ is large, this mass density is independent of $\lint$ and does not affect the stellar kinematics\footnote{We used $\kappa_{\rm composite}$ for consistency with Sec. \ref{subsec:stronglensing}, but the results presented here are valid irrespective of the parameterisation of the lens mass profile.}. %Therefore, when $r_{\rm c}$ is large, the predicted kinematics from the model is independent of $\lint$ and the $\kappa_{\rm core}$ acts like an \textit{external} mass sheet that does not affect the kinematics.

% \akineditnew{Note: I think for illustration purposes, we could adopt e.g. a PIEMD profile for $\kappa_{composite}$, for which we analytically know what the deprojected 3D density distribution looks like. In this way, we could write down Eq. C5 analytically, instead of having the "Deproj[$kappa_{composite}$]" leftovers. In essence, "Deproj[$kappa_{composite}$]" simply describes our ignorance of an analytical formula for the deprojected mass density, which we therefore perform by means of an MGE. But this is not really necessary here.}

% \comment{Sherry}{I would rather not use a PIEMD for $\kappa_{composite}$ as I think it would add further confusion to what we are doing.  I would prefer to stick to a description that matches exactly what we do, and have now added a sentence above to refer the Deproj to the MGE.  If readers have followed Section 3.2 (those who would bother to read the Appendix likely would have followed Section 3.2), they should understand the meaning of Deproj. }

\end{document}